# A Practical Approach to the Design of an S-Band Image-Rejecting Dual-Conversion Super-Heterodyne RF Chain of a Receiver Considering Spur Signals

**Seyed Mohammad Amin Shirinbayan *, Gholamreza Moradi****

**Abstract** – This paper presents a typical design of the RF section of a radar receiver, the chain within a superheterodyne dual-conversion architecture. A significant challenge in this framework is the occurrence of spur signals, which negatively impact the dynamic range of the RF chain. When addressing this issue, the paper introduces an innovative approach to mitigate (or even wipe out) these undesired effects, utilizing two mutually verifying MATLAB codes. These codes have been tested with two distinct commercial mixers and could be applied to any superheterodyne configuration with various mixers. The presented method makes the Spurious-Free Dynamic Range (SFDR) of the chain the least different from the dynamic range of the chain. Also, the selection of other components gets optimized to align with spurious signals consideration, with explanations provided for these choices. Moreover, two filters of the RF chain, the second and the third, have been designed to reduce implementation costs. Various Microwave software and full-wave analyses were employed for detailed design and analysis, with the results compared to evaluate their performance.

## 1. Introduction

Image rejection poses a significant challenge in the design of RF sections for radar receivers. Two primary approaches are employed to handle the issue of the image signal. The first method utilizes exclusive image-rejection blocks known as Hartley and Weaver architectures [1]. The second strategy capitalizes on intermediate frequencies, facilitating filtering within the intermediate frequency (IF) ranges. The latter method benefits from the substantial spectral distance between the RF signal and the image, simplifying image filtering and eliminating the need for exclusive image-rejection blocks.

Recent literature has explored various methodologies, effectively addressing the issue of image rejection. For instance, [2] examines a multiband double image rejection transmitter (DIRT) that demonstrates insensitivity to I/Q gain and phase mismatches, successfully managing image rejection across a broad frequency range (11–15 GHz). This DIRT architecture highlights the advantages of employing intermediate frequencies to mitigate image signal interference, reducing complexity and cost in RF circuit design. Concurrently, [3] introduces a tunable ultra-wideband superheterodyne RF receiver that achieves impressive image rejection levels exceeding 29 dB. This design incorporates multiple receiving channels to facilitate adjustable frequency bands, demonstrating the effectiveness of superheterodyne architectures in managing image rejection without extensive filtering.

Furthermore, [4] presents the design of a 6 GHz band RF receiver that utilizes classical superheterodyne techniques to achieve I/Q image suppression exceeding 40 dB. This design fulfills stringent noise figure requirements. Also, it optimizes dynamic range through meticulous component selection, emphasizing the necessity of minimizing spurious signals while ensuring robust image rejection capabilities. Also, [5] details a Hartley image-reject receiver that integrates a fractional-N PLL synthesizer, improving image rejection without relying on traditional filter banks and showcasing an approach to spurious signal mitigation via adaptive amplitude compensation.

Despite the mentioned works, several issues warrant attention. Firstly, image-rejector blocks are particularly suited for scenarios without multiple down-conversions or when image signal filtering is complicated due to a narrow spectral distance between the image and the desired signals. Furthermore, the calibration requirements of the image-rejector circuits can complicate the overall receiver design. Additionally, while leveraging intermediate frequency advantages and multiple filter banks can enhance image signal suppression beyond previously documented, the mentioned works tend to increase the overall complexity of the components comprised in the RF chain.

* Dept. of Electrical Engineering, Amirkabir Univerity of Technology, Iran. (m.a.shirinbayan@aut.ac.ir)

** Dept. of Electrical Engineering, Amirkabir Univerity of Technology, Iran. (ghmoradi@aut.ac.ir)

A significant drawback of employing multiple filter banks is the heightened presence of spurious signals due to the numerous mixers within the RF chain, which can severely limit the overall dynamic range of the system. This paper aims to address these challenges, specifically focusing on spur signal mitigation within the RF chain of a superheterodyne receiver utilizing multiple filter banks for image rejection. With the assistance of two MATLAB-based codes analyzing the issue of spurious signal and component selection, the proposed design seeks to improve overall system performance while minimizing the complexity of the components comprised in the RF chain and cost. To this end, the study in [6] employs a superheterodyne architecture to manage the complexities associated with multiple down-conversions, underscoring the necessity of effectively filtering out spur signals. However, a more comprehensive circuit analysis for such a designed RF chain still seems demanding. Therefore, in this work, proposing a method for identifying frequency regions devoid of spurious signals, the authors emphasize the significance of frequency planning in achieving optimal dynamic range performance. Consequently, when selecting the appropriate components of the RF chain alongside designing an IF filter, the whole RF chain will be determined. Finally, simulations of the entire RF chain will be conducted using the CASCADE software and a Link Budget analysis via the ADS Keysight, delivering comprehensive circuit analysis for a designed RF chain.

## 2. MATLAB-based Approach Dealing with Spurious Signal Consideration in an RF Chain

As previously noted, one method to tackle the issue of the image signal is to utilize Hartley and Weaver circuits in the RF chain's architectures. However, both circuits have non-ideality, making the utilization difficult in applications that require image rejection of more than 40 dB. Such circuits demand calibration techniques that automatically make the whole RF chain more complicated [1]. Also, the presence of a power divider, phase shifter, and other elements attached to the circuit, in addition to increasing the complexity of the RF chain design, can contribute to higher implementation costs. One of the prevalent solutions for this problem is to conduct several down-conversions of frequency using superheterodyne architectures. Through the architectures, image rejection is done by filtering the signal at higher frequencies, exploiting the advantages of intermediate frequencies [7]. However, a substantial challenge associated with these architectures is that using multiple mixers results in the presence of spur signals, significantly constraining the overall dynamic range of the RF chain [7] [8]. An effective strategy to mitigate or surpass the detrimental impact of spur signals on dynamic range involves identifying frequency regions devoid of these unwanted signals [7] [9]. Appropriate selection of the IF and LO signal spectra can significantly minimize the negative impact of spur signals, provided that the spectra of these spur signals do not overlap with those of the desired signals. Such a reduction can be achieved through effective frequency filtering. Thus, addressing the issue of spur signals relies on developing an effective frequency plan. To effectively resolve this issue, it is significant to carefully select the first, second, and subsequent LO and IF frequencies to ensure that spur signals do not interfere with the desired signal spectrum. To this end, the following subsections of this paper present an appropriate scheme [7].

Before discussing the presented approach, it is worth noticing that the superheterodyne architecture presented here is assumed to have two down-conversions. Also, the RF bandwidth related to the input bandwidth of the first mixer is set to 400 MHz, ranging from 2700 MHz to 3100 MHz, with a central frequency of 2900 MHz. The output bandwidth of the second mixer, just before the analog-to-digital converter (ADC), is assumed to be 5 MHz, while the targeted dynamic range is established at 70 dB.

### 2.1 Finding Spurious-Free Frequency Regions with the MATLAB Software Toolbox

Establishing an optimal frequency range for the LO and IF signals—the appropriate frequency plan—is crucial for mitigating the detrimental effects of spurious signals. Such a plan ensures that the spectra of the spurious signals, restricting the dynamic range, are not positioned in proximity to that of the desired signals at the output of the mixers. In this paper, we present a fitting method that addresses this challenge. The proposed method is applied to two specific mixers, the MCA-60+ [10] and ADE-MH35+ [11], chosen for the RF chain in a typical radar receiver in this study. The selection of these mixers from the Mini-Circuits company was primarily based on their excellent performance within the specified frequency range. Also, both mixers are Surface Mount Devices (SMDs) and exhibit relatively high Third Order Input Intercept Points (IIP3s).

To clarify the approach in tackling the issue of spurious signals, initially assume the MCA1-60+ mixer as the first frequency down-converter. In the datasheet of each mixer, it is customary to find a harmonic table that displays the power levels of each spurious harmonic present at the output of the mixer compared to the power of the main harmonic, which represents the desired signal at the output of the mixer [10]. Furthermore, the datasheet also provides details related to the test conditions under which the data in the table were obtained. Table 1 presents an illustrative example of such a table for the MCA1-60+ mixer.

**Table 1.** The spurious table for the MCA1-60+ mixer [10].

| RF HARMONICS ORDER | (-dBm) | (-dBc) | | | | | | | | | | |
|---|---|---|---|---|---|---|---|---|---|---|---|---|
| 0 | - | - | +13 | 20 | 20 | 39 | 15 | 45 | --- | --- | --- | --- |
| 1 | - | 10 | +0 | 28 | 20 | 42 | 35 | 50 | 45 | --- | --- | --- |
| 2 | >90 | 52 | 46 | 51 | 46 | 53 | 66 | 53 | 48 | 67 | --- | --- |
| 3 | >90 | 65 | 59 | >69 | 63 | 62 | >69 | >69 | >69 | >69 | >69 | --- |
| 4 | >90 | >69 | >69 | >69 | >69 | >69 | >69 | >69 | >69 | >69 | >69 | >69 |
| 5 | >90 | >69 | >69 | >69 | >69 | >69 | >69 | >69 | >69 | >69 | >69 | >69 |
| 6 | >90 | >69 | >69 | >69 | >69 | >69 | >69 | >69 | >69 | >69 | >69 | >69 |
| 7 | --- | --- | >69 | >69 | >69 | >69 | >69 | >69 | >69 | >69 | >69 | >69 |
| 8 | --- | --- | --- | >69 | >69 | >69 | >69 | >69 | >69 | >69 | >69 | >69 |
| 9 | --- | --- | --- | --- | >69 | 65 | >69 | >69 | >69 | >69 | >69 | >69 |
| 10 | --- | --- | --- | --- | --- | >69 | >69 | >69 | >69 | >69 | >69 | >69 |
| | RF CAL | 0 | 1 | 2 | 3 | 4 | 5 | 6 | 7 | 8 | 9 | 10 |

Test conditions: RF IN: 3800 MHz; -14.00 dBm. LO IN: 3830 MHz; +7.00 dBm IF OUT: 30 MHz; -20.98 dBm

In Table 1, the number of each vertical and horizontal row corresponds to the coefficients of each of the signals in the output of the mixer [9]. The red square at the intersection of the first vertical and first horizontal rows indicates the desired signal generated by the frequency difference between the LO frequency and that of RF ($|f_{RF} - f_{LO}|$, $M=1, N=1$). In other words, the position of each square in the table corresponds to the coefficients in the expression $|Mf_{RF} - Nf_{LO}|$. The intersection of horizontal row number M and vertical row number N corresponds to the spurious signal at the mixer's output with a frequency of $|Mf_{RF} - Nf_{LO}|$.

Within each square, the value represents the discrepancy in power level between the spurious signal associated with that square and the desired output power level. For instance, in the table, when it states "46" with a frequency of $|2f_{RF} - f_{LO}|$ (that is $M=2, N=1$), it indicates that the power of the mixer's output at this specific frequency is 46 dB lower than the power of the desired signal located in the red square.

Also, in Table 1, the green squares represent spurious signals, having a power level of more than 70 dB lower than the desired signal (non-impact spur). These spurious signals, categorized as non-destructive spurs, do not limit the dynamic range. The yellow squares indicate spurious signals with an output power between 50 and 70 dB lower than the desired signal (moderate spur). The blue squares indicate spurious signals with an output power of at most 50 dB lower than the desired signal (critical spur). The yellow and blue squares represent unwanted spurious signals and should be avoided in proximity to the selected IF range. Further details regarding this classification will be addressed in section 2.2.

The frequency plan determines whether spurious signals are present at the mixer's output spectra. If these signals were at the mixer's output spectrum and their power level was not sufficiently lower than the desired signal, the output signal could be distorted. Such distortion would be visible in the time domain. Thus, avoiding these detrimental signals within the passband regions would be crucial. If they existed within the passband, their power should be significantly lower than that of the desired signal to ensure that the resulting distortion is negligible.

The RF section of a typical S-band receiver often includes one or more non-linear components. These components could introduce distortion to the desired signal, partially restricting the dynamic range. To ensure that the non-linear mixer component does not impede the dynamic range, the power of the resulting spurious signals should be lower than that of the desired signal. This power difference between the desired and the spurious signal should exceed the dynamic range to avoid introducing any distortion. For example, if the desired dynamic range is 70 dB, the power of the spurious signals within the passband should be at least 70 dB lower than that of the desired signal. Ideally, it is preferable to establish a frequency plan that prevents the placement of spurious signals within the passband altogether.

Utilizing the method described in this article, it becomes feasible to establish a frequency plan for each mixer that prevents the presence of detrimental spurious signals in the passband, which would otherwise limit the dynamic range. The methodology utilized here for determining the frequency plan relies on two MATLAB codes that mutually verify each other. The first code utilizes purpose-built functions within the MATLAB toolbox. The code operates by getting input parameters from the user, including the RF signal's center frequency, RF bandwidth, IF bandwidth, the minimum acceptable power difference between the spurious signal and the desired signal, and the spurious table associated with the particular mixer used. The code generates output that determines regions where placing the IF center would avoid unwanted spurious signals within the designated IF bandwidth. In other words, this code assists the user in identifying spurious-free areas. Figure 1 provides an example of the MATLAB code's output for the MCA1-60+ mixer.

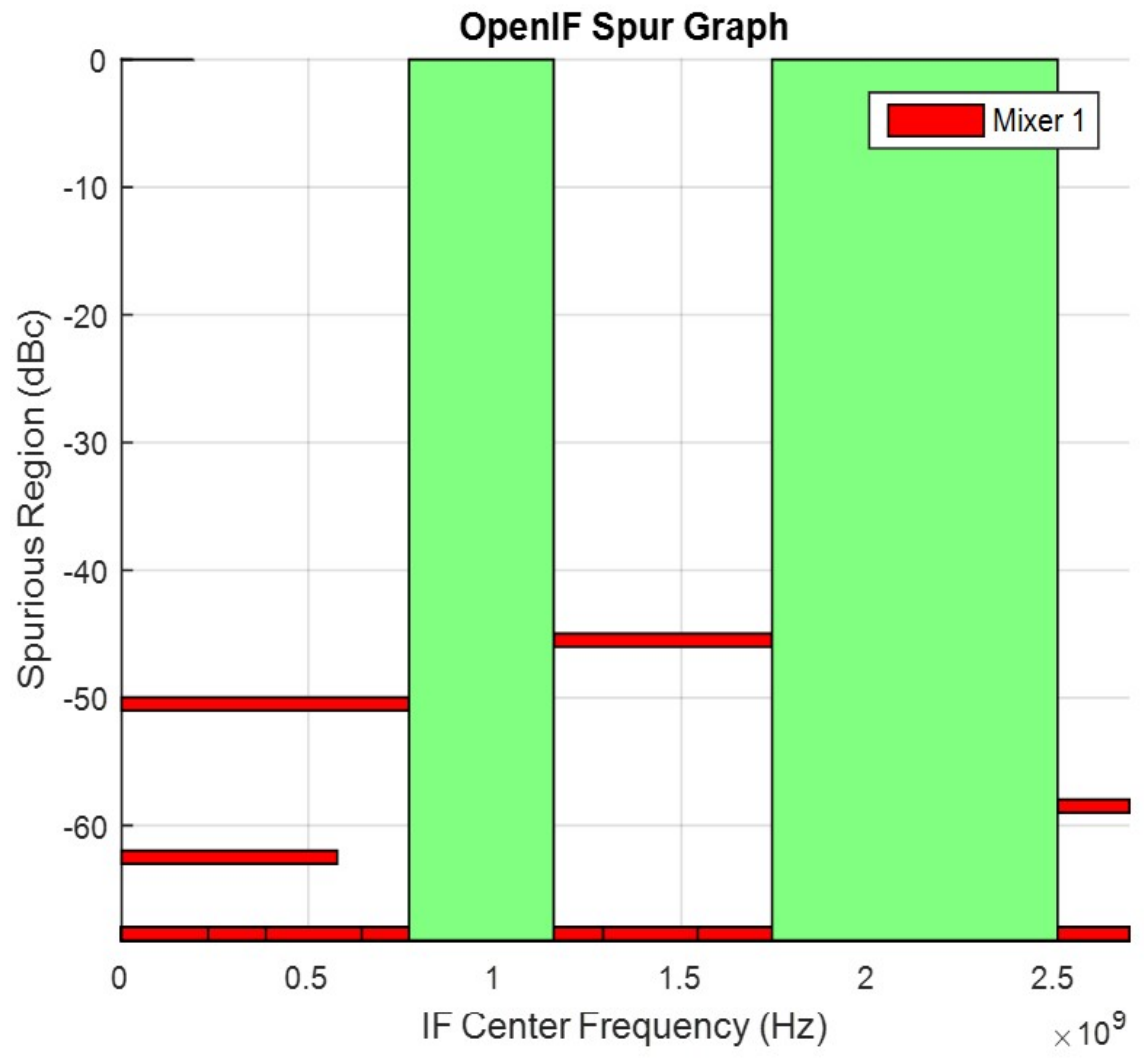

**Fig. 1.** Spurious-free regions for the MCA1-60+ mixer.

Figure 1 is the output of the MATLAB toolbox codes using the spurious table provided in the MCA1-60+ mixer datasheet. The center frequency of the input signal's spectrum is 2.9 GHz, with an input bandwidth of 400 MHz and an output bandwidth of 30 MHz. The down-conversion type considered here is high-side injected due to the assumption made in the MCA1-60+ datasheet, mentioned in Table 1. Consequently, the LO frequency should be higher than the RF frequency. Figure 1 displays the potential location of the IF signal's center frequency. The green areas represent regions where the spurious signals do not limit the dynamic range when the IF signal's frequency center is there. The red lines indicate the relative power level of the spurious signals within the IF range. The vertical axis illustrates the power level of these signals relative to the desired signal. Clicking on each red line in Figure 2 reveals the corresponding coefficients for the spurious signal, while clicking on the green areas provides the desired frequency range for placing the IF signal's center frequency.

Given the green regions depicted in Figure 2, there is a possibility of spur signals being present. However, if such signals exist, their magnitude is significantly lower than that of the desired IF signal, ensuring they do not limit the dynamic range. Therefore, their distortion effect is trivial and could be waived. Also, the MATLAB toolbox code facilitates the determination of the minimum power of permitted spurious signals in the spurious-free regions, which is controlled by the parameter named **`h1.SpurFloor`** in the **`OpenIF`** function defined in the MATLAB Toolbox.

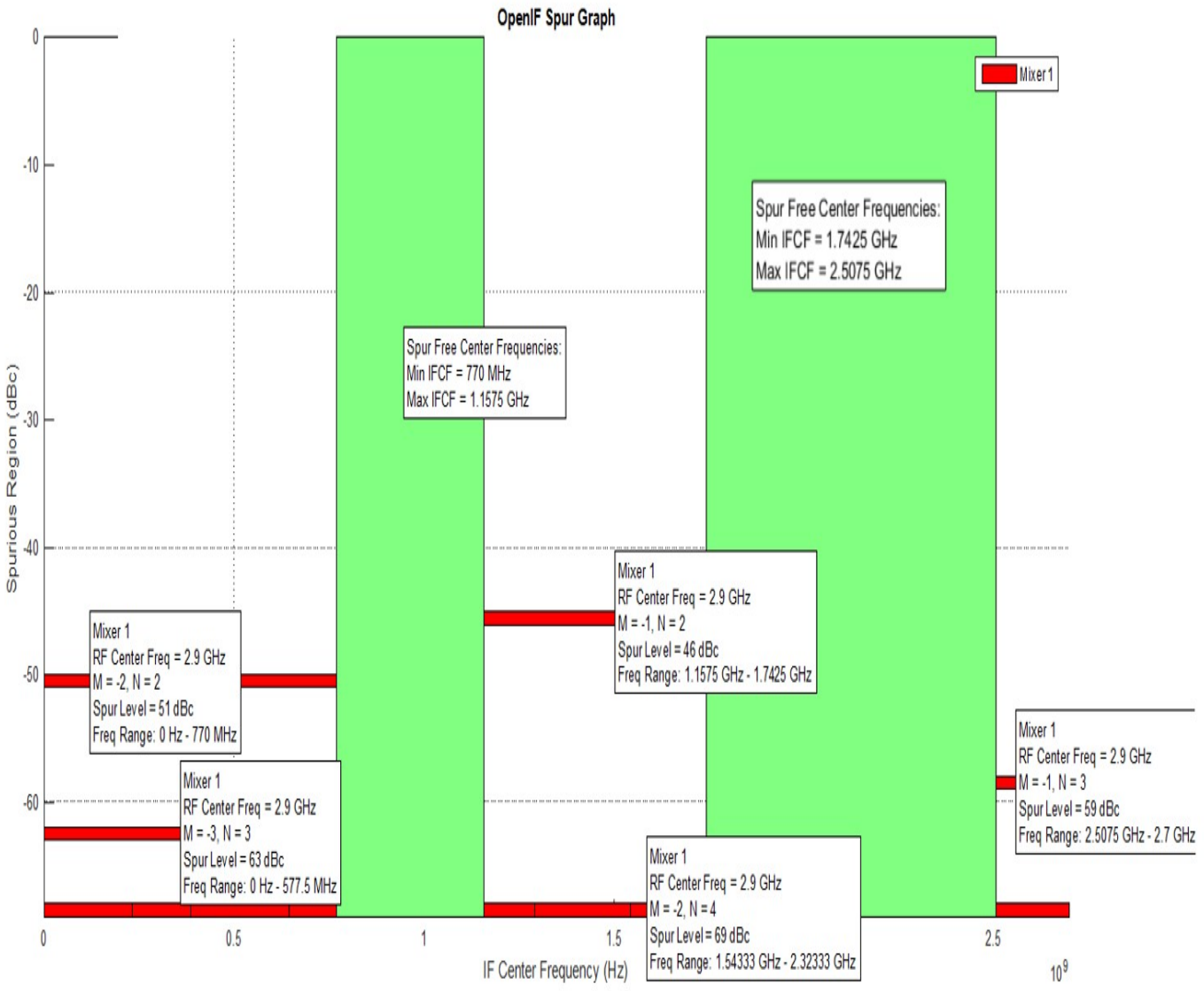

**Fig. 2.** Illustration of spurious-free regions and areas of potential spurious signals with corresponding coefficients for the MCA1-60+ mixer

Based on the output of Figure 2, the IF bandwidth central frequency for MCA1-60+ was opted for 1800 MHz. As illustrated in Figure 2, the center of the intermediate frequency falls within the spurious-free region, guaranteeing an unhindered dynamic range. Consequently, considering the high-side injection down-conversion, the LO frequency of the first mixer is set to 4700 MHz ( $=1800MHz + 2900MHz$ ). Also, assuming a 30MHz output bandwidth for the first mixer of the RF chain, the first IF's bandwidth spans from 1785 MHz to 1815 MHz.

### 2.2 Verification of Spurious-Free Regions Using the MATLAB Software Toolbox

Following section 2.1, a method was proposed to identify spurious-free regions using the MATLAB toolbox. The next step is to verify whether these areas involve confirming the absence of these unwanted signals within the identified areas. Typically, a diagram is utilized to validate the frequency plan selected for receivers and transmitters, as shown in Figure 3 [8].

Figure 3 offers a visual representation where the normalized output frequency of the mixer is on the vertical axis, and the normalized input frequency is on the horizontal axis. There is an assumption that the frequency of the RF signal is lower than the frequency of the LO signal. Thus, the diagram implies a high-side injection in the mixer. Due to such an assumption, the LO frequency is denoted as "H" and the RF frequency as "L"( $H = f_{LO}$ , $L = f_{RF}$ ). The sweeping lines on the graph depict the frequencies observed at the mixer's output. Each point along a particular line in the diagram corresponds to a signal's frequency (which can be either a spur signal or the desired signal) at the output of the mixer, along with the normalized frequency at the input of the mixer, corresponding to that particular frequency. The squares on the graph indicate the free-spurious regions, making them appropriate spans for selecting the IF bandwidth central frequency.

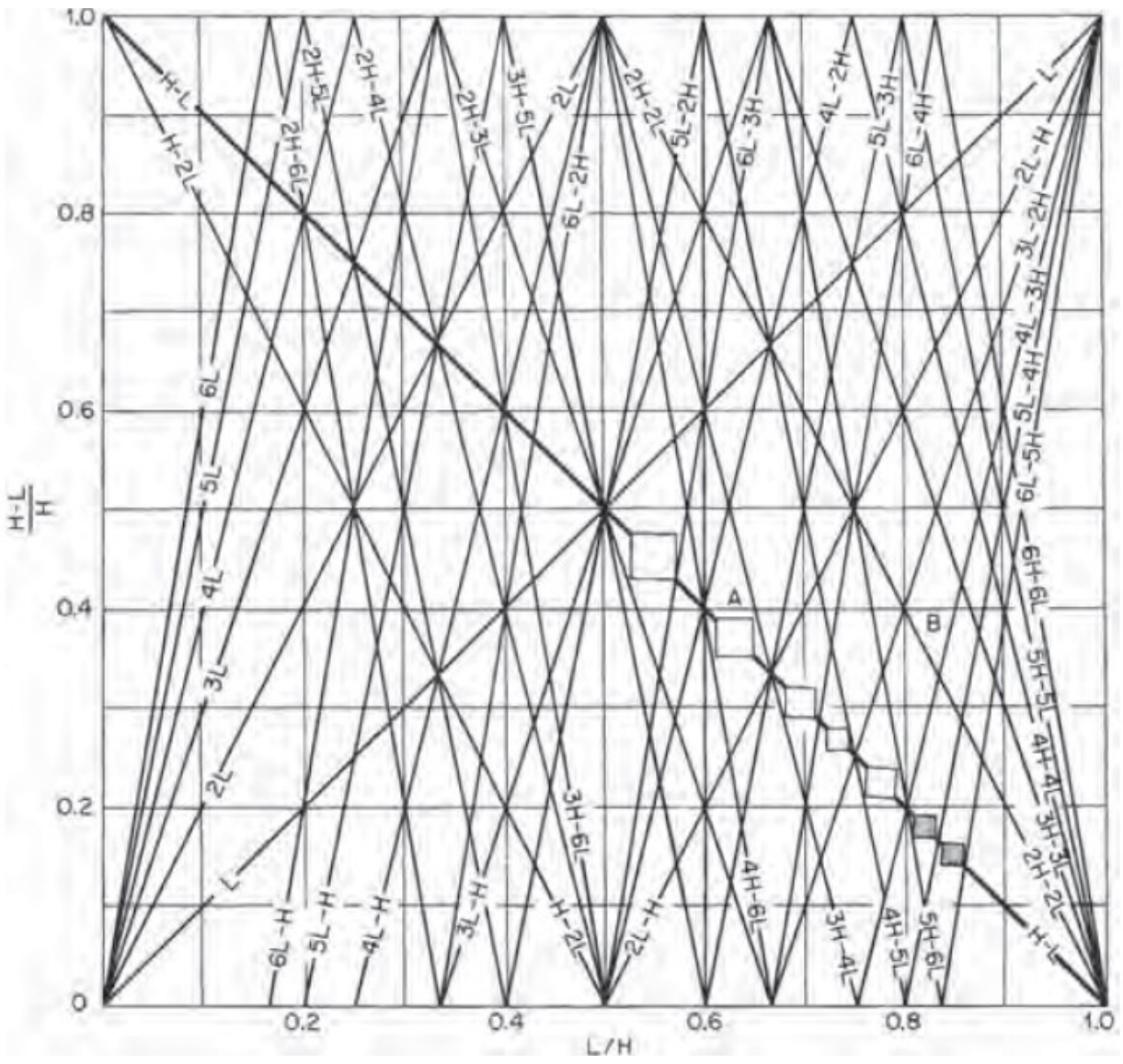

**Fig. 3.** Spur diagram for frequency down-conversion in a mixer (LO frequency: H, RF frequency: L) [8]

To verify the spurious regions determined using the MATLAB toolbox, we utilize the methodology presented in Figure 3. To this end, a manual code was developed in MATLAB using the spurious table provided in the MCA1-60+ mixer datasheet. The resulting diagram, Figure 4, similar to the one depicted in Figure 3, is generated as the output of this manual code.

Figure 4 presents a graphical representation where the red line represents the desired intermediate frequencies at the mixer output. These frequencies are calculated by subtracting the LO signal frequency from the RF signal frequencies. Spurious signals are also depicted in Figure 4 using three distinct colors. The choice of colors is based on the power difference between any of these spurious signals compared to the desired signal, as indicated in Table 1.

In section 2.1, based on the spur-free regions depicted in Figure 2, the output central frequency of the first selected mixer of the RF chain (MCA1-60+) was designated 1800 MHz. Furthermore, a frequency span without spurious signals was identified within the frequency range of 1785 MHz to 1815 MHz, which can be utilized as the first IF bandwidth. Upon examining the output of the manual MATLAB code employed in this section, it becomes evident that the area enclosed within the black rectangle, including the red line passing through its two vertices, in Figure 4 overlaps with the spurious-free region previously selected.

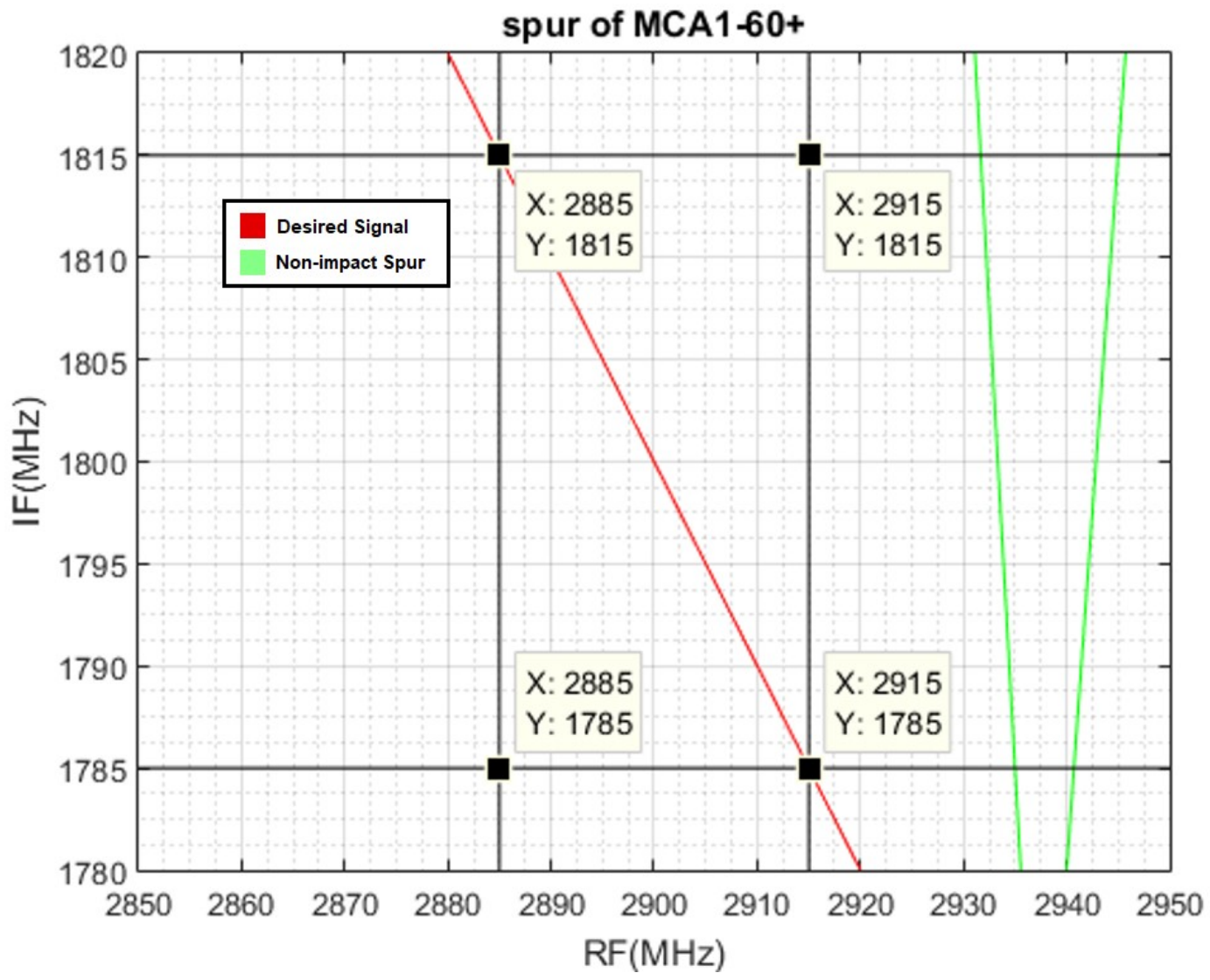

**Fig. 4.** Output of the manual MATLAB code for the MCA1-60+ mixer indicating optimal range for the first intermediate frequency (IF1) within a black rectangle.

It is worth noting that the MATLAB toolbox code output did not explicitly indicate whether this area was entirely free of spurious signals or contained non-destructive spurs. However, using the manual code presented in this section, it can be perceived that this area lacks spurious signals, validating the appropriateness of the selected IF region. In essence, this implies that by choosing the frequency range of 1785 MHz to 1815 MHz for the MCA1-60+ mixer's output, the spurious signals generated by the mixer do not impose any constraints on the dynamic range, even if the desired dynamic range exceeds 70 dB.

Both MATLAB codes developed in this paper were based on the spur table from the MCA1-60+ mixer datasheet. However, the MATLAB toolbox code and the manual code can jointly be adapted to determine the frequency plan of any other mixer by modifying the corresponding spur table.

In the receiver design discussed further in this paper, section 3, two mixers are employed for the dual down-conversion. For the second down-conversion, the ADE-MH35+ mixer could be an appropriate option. The ADE-MH35+, like the first mixer of the RF chain (MCA1-60+), is manufactured by Mini-Circuits Company and offers excellent performance within the specified frequency range.

As discussed earlier, the center frequency of the first IF is set at 1800 MHz, accompanied by a bandwidth of 30 MHz. The approach used to determine the frequency for the second IF, located at the output of the ADE-MH35+ mixer, is the same as the method applied for the first IF. This approach first involves using the MATLAB toolbox code to determine an appropriate spurious-free region and then verifying the identified region using the manual MATLAB code exclusively developed for the spurious table in the ADE-MH35+ mixer's datasheet [11]. The spurious table for the ADE-MH35+ mixer can be seen in Table 2, while the MATLAB toolbox's output for the ADE-MH35+ mixer is depicted in Figure 5.

According to Table 2 and the determined spurious-free regions, the center frequency of the output bandwidth for the second mixer can be regarded as 60 MHz. Consequently, the second local oscillator's frequency is 1860 MHz, as the second mixer also down-converts its input signal with high-side injection. Furthermore, given that the assumed output bandwidth of the second mixer is 5 MHz, the bandwidth of the second IF ranges from 57.5 to 62.5 MHz.

**Table 2.** The spurious table for the ADE-MH35+ mixer [11].

| RF HARMONICS ORDER | (-dBm) | (-dBc) | | | | | | | | | | |
|---|---|---|---|---|---|---|---|---|---|---|---|---|
| 0 | - | - | +6 | 5 | 8 | 16 | 18 | 26 | 22 | 35 | 31 | 51 |
| 1 | - | 17 | +0 | 29 | 16 | 32 | 19 | 35 | 30 | 39 | 35 | 40 |
| 2 | 85 | 47 | 49 | 43 | 70 | 46 | 52 | 46 | 49 | 58 | 52 | 65 |
| 3 | >100 | 70 | 58 | 84 | 61 | >86 | 63 | 74 | 62 | 71 | 65 | 70 |
| 4 | >100 | >86 | >86 | 81 | >86 | 76 | >86 | 79 | >86 | 77 | >86 | >86 |
| 5 | >100 | >86 | >86 | >86 | >86 | >86 | >86 | >86 | >86 | >86 | >86 | >86 |
| 6 | >100 | >86 | >86 | >86 | >86 | >86 | >86 | >86 | >86 | >86 | >86 | >86 |
| 7 | >100 | >86 | >86 | >86 | >86 | >86 | >86 | >86 | >86 | >86 | >86 | >86 |
| 8 | >100 | >86 | >86 | >86 | >86 | >86 | >86 | >86 | >86 | >86 | >86 | >86 |
| 9 | >100 | >86 | >86 | >86 | >86 | >86 | >86 | >86 | >86 | >86 | >86 | >86 |
| 10 | >100 | >86 | >86 | >86 | >86 | >86 | >86 | >86 | >86 | >86 | >86 | >86 |
| LO HARMONICS ORDER | RF CAL | 0 | 1 | 2 | 3 | 4 | 5 | 6 | 7 | 8 | 9 | 10 |

Test conditions:
RF IN: 1750.1 MHz; -6.00 dBm.
LO IN: 1780.01 MHz; +13.00 dBm
IF OUT: 29.91 MHz; -14.46 dBm

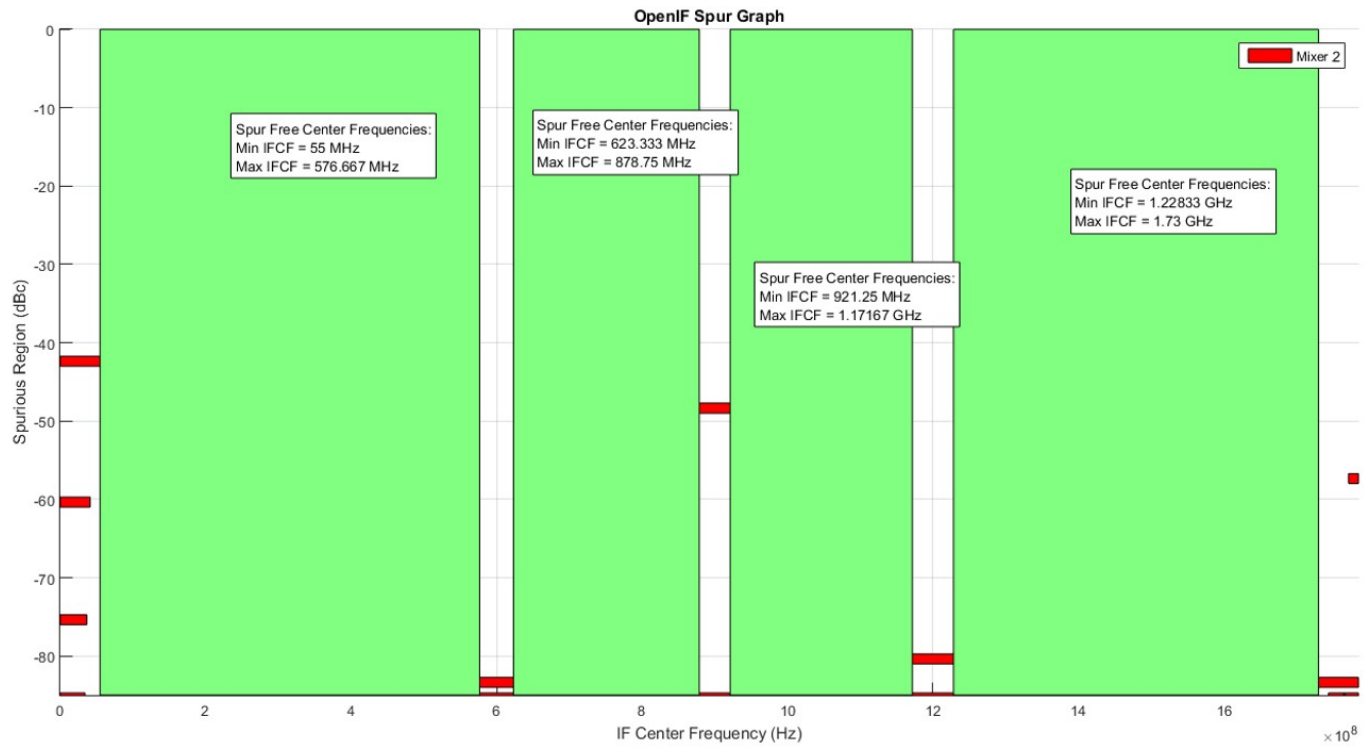

**Fig. 5.** Depiction of spurious-free regions and areas of potential spurious signals with corresponding coefficients for the ADE-MH35+ mixer.

Interestingly, the MATLAB toolbox's output, Figure 5, shows several broad spurious-free regions (the green regions) below the 1800 MHz frequency. The reason for the abundance of these spurious-free regions is that the input bandwidth of the second mixer (30 MHz) is significantly narrower compared to the input bandwidth of the first mixer (400 MHz). This low bandwidth massively prevents signals with spurious frequencies from going through the second mixer's input. As a result, more spurious-free regions can be identified for the second mixer aligned with the result conducted by examining the manual code. However, Figure 5 might not sufficiently convey this issue. For a more comprehensive view, the outputs of the manual code, developed using MATLAB, could be considered as the graphs presented in Figures 6-9.

In Figures 6 and 7, the region between the two black parallel lines denotes the selected frequency range for the output of the second mixer. The following figures, Figures 8 and 9, provide a more detailed view of such a frequency span. The selected frequency span between the two black parallel lines, depicted in Figures 8 and 9, contains only the desired frequencies without interference from spurious signals.

The consistency between the manual code output and the MATLAB toolbox code output is evident. For example, the frequencies of the spurious signal represented by the blue dashed line in Figures 7-9 can also be observed in the output generated by the MATLAB toolbox code in Figure 5 or Figure 10.

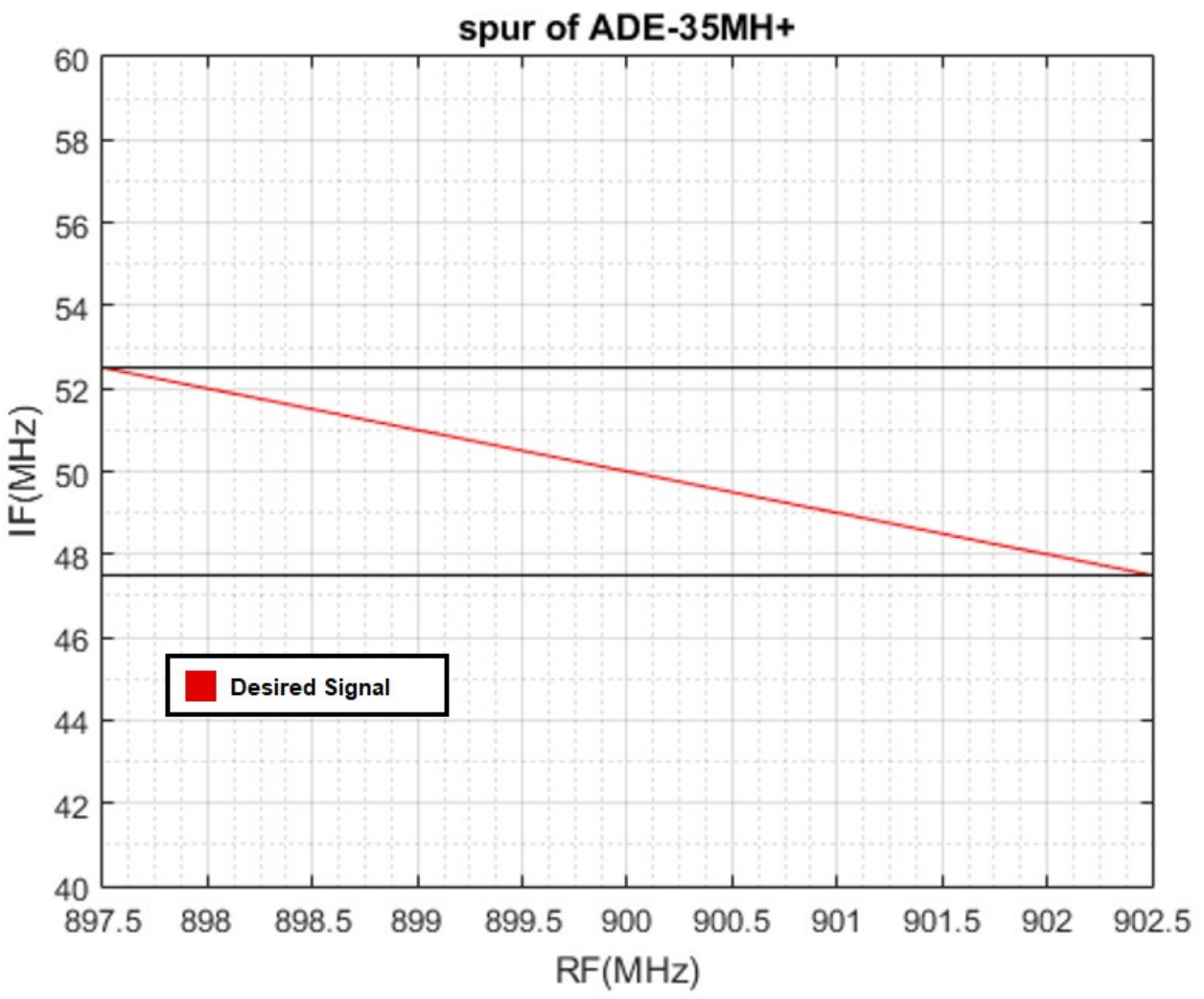

**Fig. 6.** Output of the manual MATLAB code for the ADE-MH35+ mixer indicating optimal range for the second intermediate frequency (IF2) between two parallel black lines.

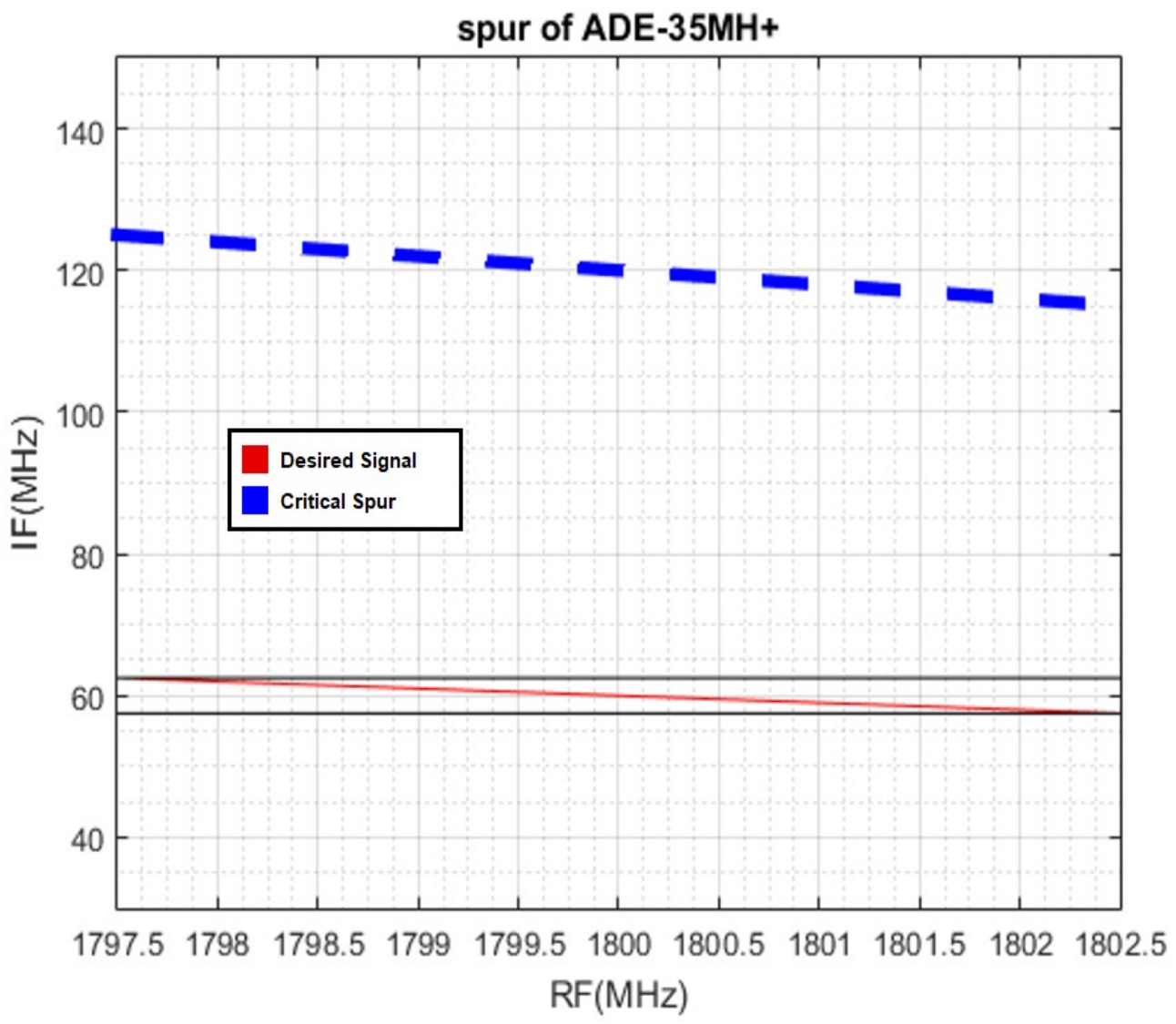

**Fig. 7.** More detailed output of the manual MATLAB code for the ADE-MH35+ mixer indicating the desired frequency range for the second IF (IF2) with the red line and disruptive spur frequency range with the blue dashed line aligned with a blue square in Table 2.

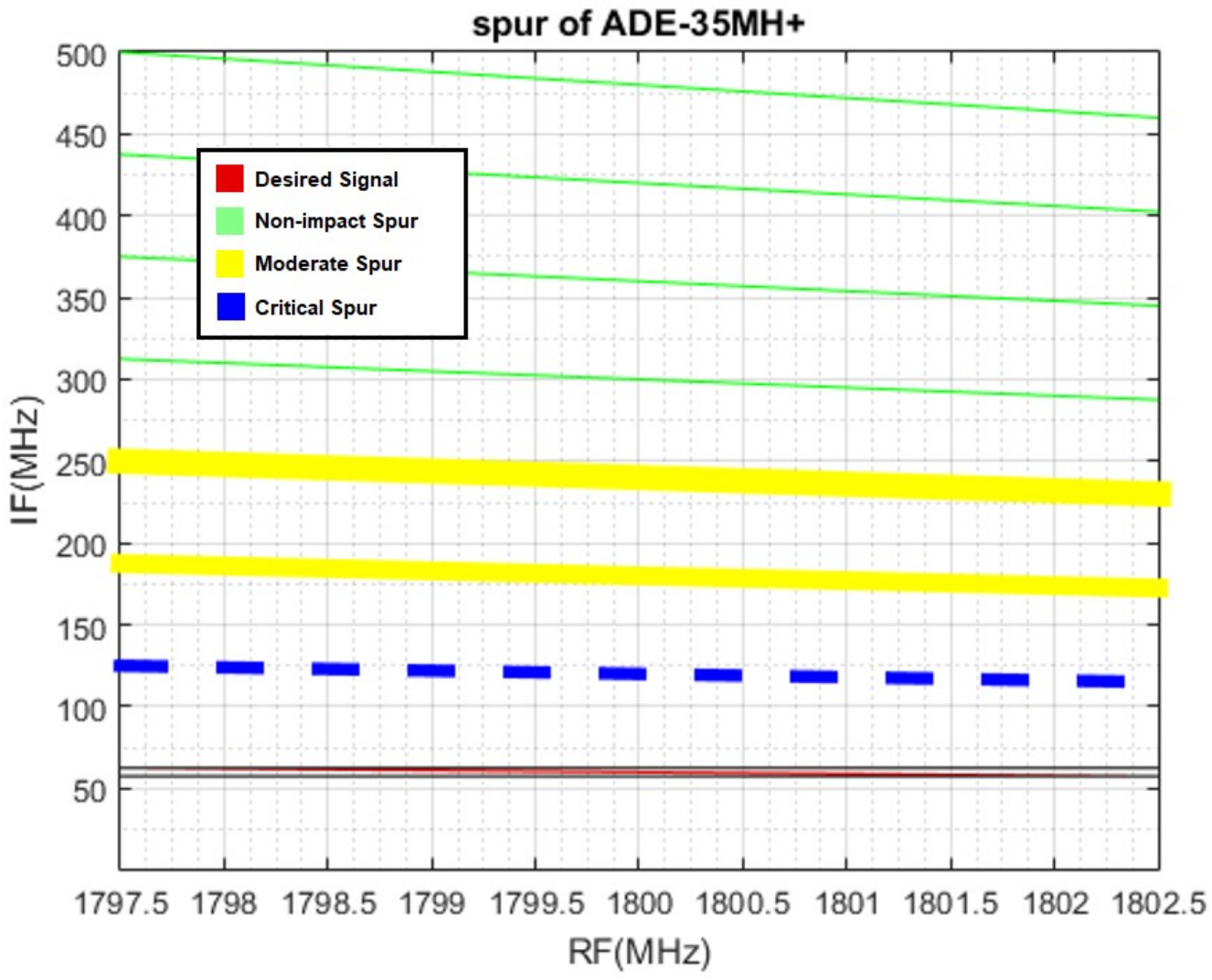

**Fig. 8.** Output presented in greater detail, with solid yellow lines and thin green lines representing additional spur frequency ranges corresponding to similar colors in Table 2.

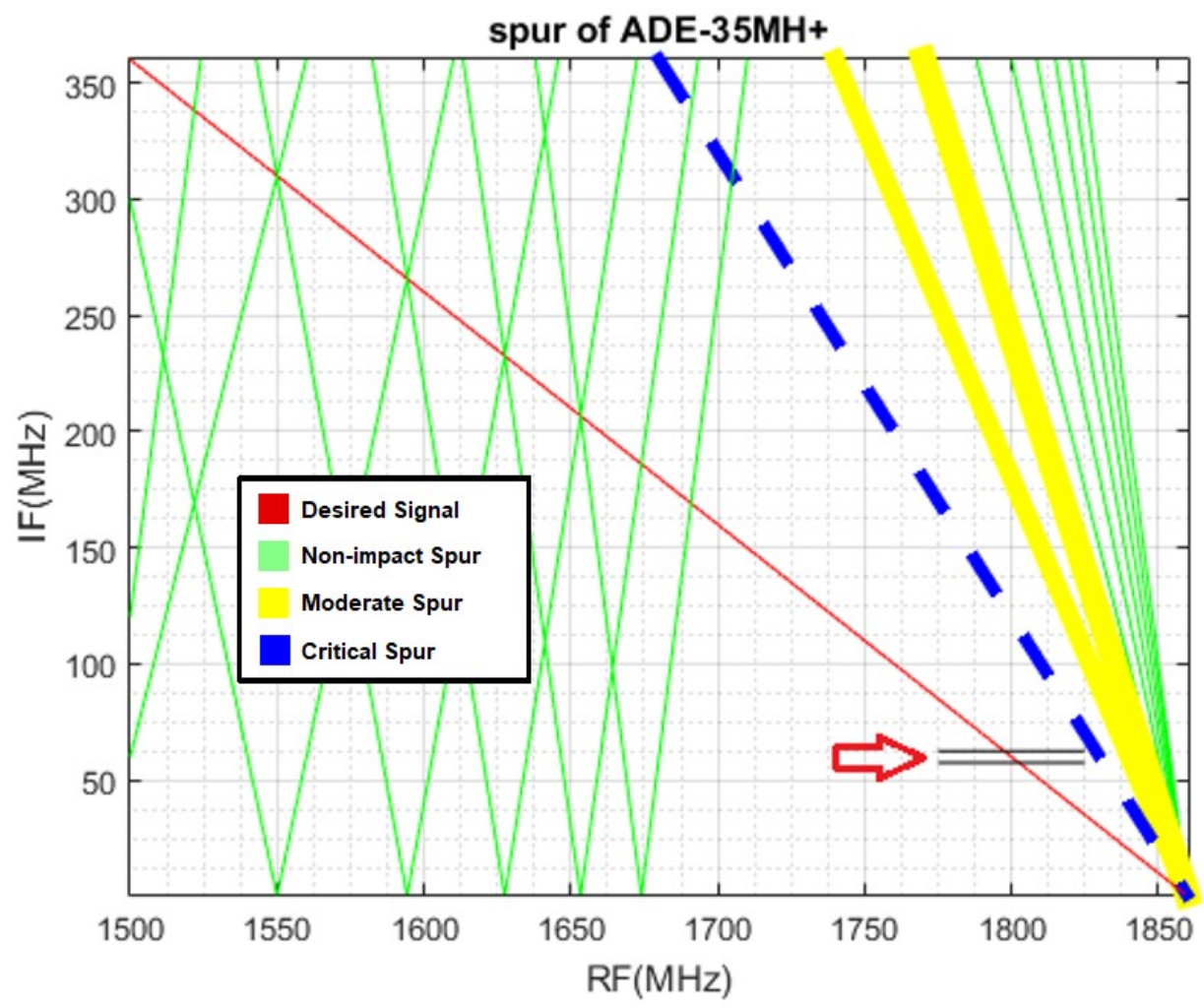

**Fig. 9.** An overview on the output of the manual MATLAB code for the ADE-MH35+ mixer in wide frequency range.

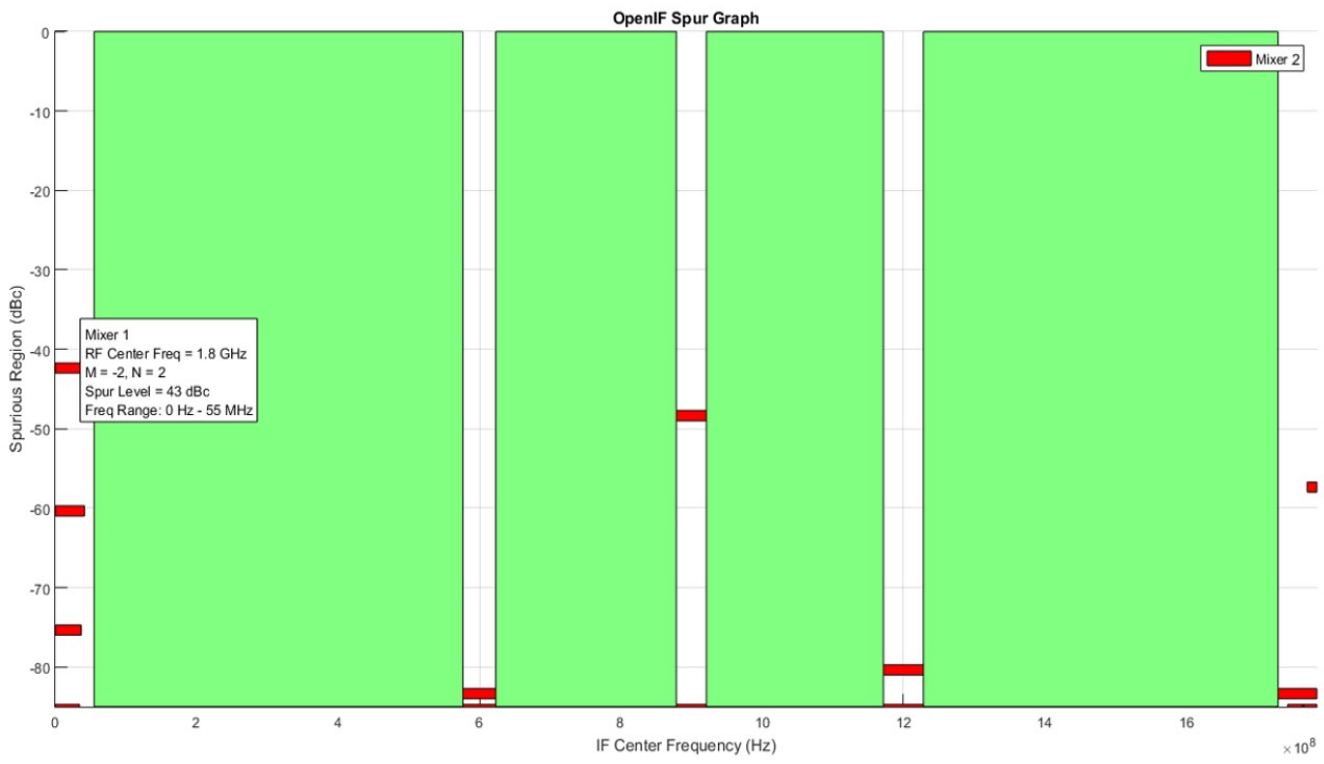

**Fig. 10.** Alternative view of Figure 5, providing additional details about the spur frequency with coefficients $M = -2$ and $N = 2$.

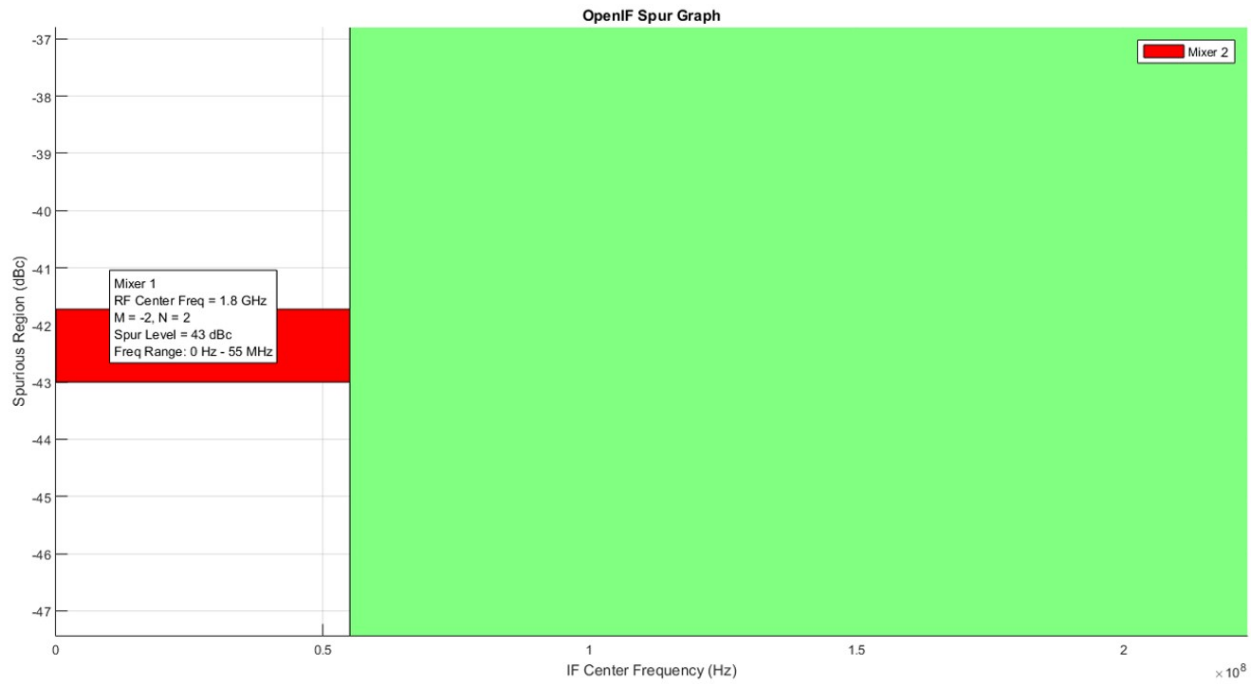

**Fig. 11.** Zoomed-in view of Figure 10

To address any doubts about the relation between the manual code output and the MATLAB toolbox code output, for instance, the blue dashed line showing the frequencies of the spurious signal in Figure 7 and the solid red line marked in Figures 10 and 11 ($|2f_{RF} - 2f_{LO}|$, $|M| = 2, N = 2$), could be examined by the information provided in Figure 12.

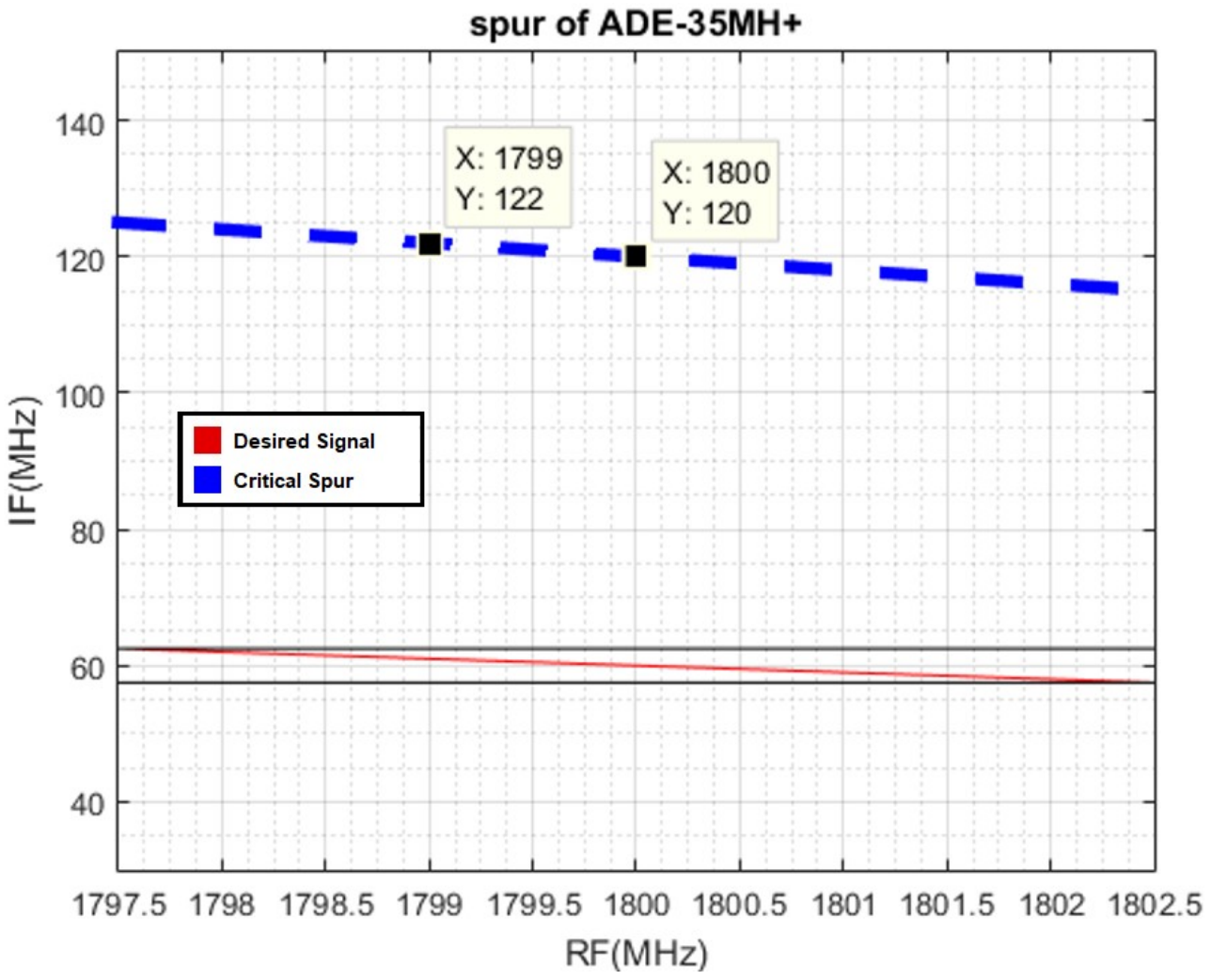

**Fig. 12.** Alternative view of Figure 7, providing additional details about the particular spur frequency.

To clarify the coefficients of the spur frequency, we could consider two typical points on the blue dashed line in Figure 12 that represent the location of the spur frequencies at the mixer's output and input. In this case, the mixer performs frequency down-conversion, where the output frequency of the mixer is the difference between the input and LO frequency. Accordingly, solving $|N \times 1799 + M \times 1860| = 122$ and $|N \times 1800 + M \times 1860| = 120$ results in $N = 2$ and $M = -2$, demonstrating that the coefficients calculated for the observed spur in the output of the manual MATLAB code in Figure 12 are consistent with the ones shown in Figures 10 and 11. Correspondingly, this cross-verification between the outputs of the two implemented codes can be extended to analyze other potential spurs.

## 3. Identifying Key Parameters for System Component Selection and Design Overview

The receiver design is based on the superheterodyne principle with dual down-conversion. The superheterodyne architecture requires careful consideration of various specifications. To determine the specifications for the RF section of the receiver, the relations and assumptions used in references [1], [7], and [8] are utilized. The receiver specifications are defined as follows in Table 3:

**Table 3.** The desired receiver specifications for design

| Parameter | Value |
|---|---|
| Gain | >50 dB |
| Dynamic Range | >70 dB; -132 dBm to -32 dBm |
| OIP1 | >15 dBm |
| OIP3 | >33 dBm |
| Noise Figure (NF) | < 5 dB |

The designed RF chain comprises three fixed-gain amplifiers, one variable-gain amplifier, two mixers, and passband filters. Figure 13 illustrates the arrangement and operational frequency for each component.

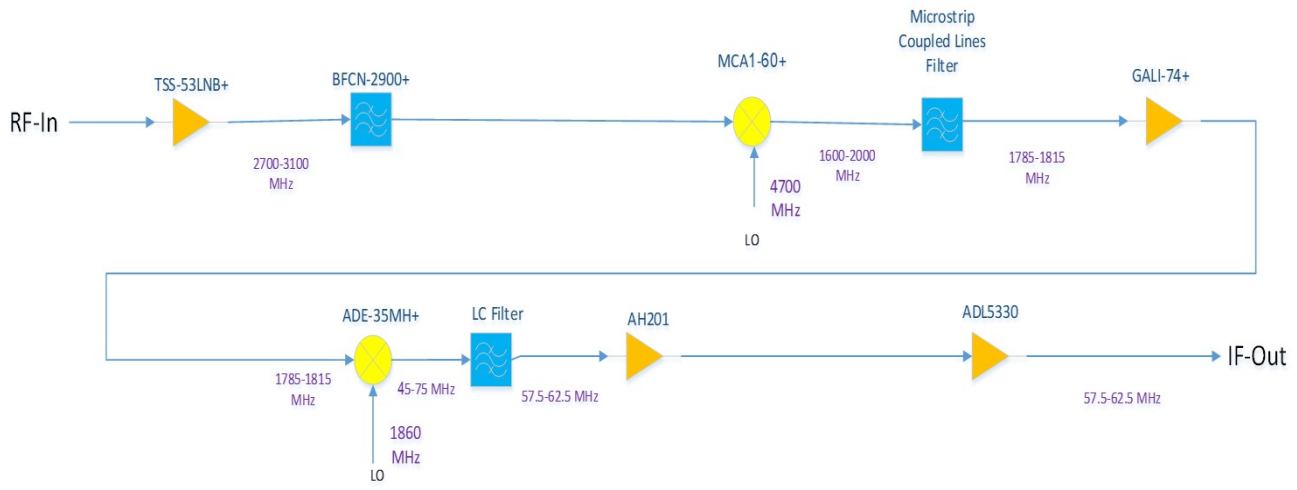

**Fig. 13.** RF chain of the receiver and the selected components

As shown in Figure 15, the two yellow circles represent the mixers, while the triangles depict the LNA, RF Amplifier, IF Amplifier, and VGA. Additionally, the light blue squares indicate the filters. In the following sections, we will explore the process of selecting these various components.

### 3.1 Mixers

The frequency plan and mixer selection for the RF chain have been designed to effectively mitigate the impact of spurious signals based on the approach outlined in section 2. Therefore, using the superheterodyne architecture with frequency dual down-conversion, an appropriate performance in image rejection and reducing the destructive effect of the spur signals could be achieved. According to section 2, the selected mixer components are the MCA1-60+ and ADE-MH35+. As mentioned, the chosen mixers are manufactured by Mini-Circuits and have demonstrated proper performance within the specified frequency range, as indicated by the high input IP3 value. Their frequency range is well-suited for the specific application requirements, Table 4 and Table 5 [10] [11].

**Table 4.** Specifications of the MCA-60+ mixer

| Model Number | LO Level (dBm) | RF Freq. Low (MHz) | RF Freq. High (MHz) | LO Freq. Low (MHz) | LO Freq. High (MHz) | IF Freq. Low (MHz) | IF Freq. High (MHz) | Conversion Loss (dB) Typ. | LO-RF Isolation (dB) Typ. | LO-IF Isolation (dB) Typ. | Input P1dB (dBm) Typ. | Input IP3 (dBm) Typ. | Subcategory |
|---|---|---|---|---|---|---|---|---|---|---|---|---|---|
| MCA1-42MH+ | 13 | 1000 | 4200 | 1000 | 4200 | DC | 1500 | 6.2 | 32 | 20 | 9 | 16 | Double Balanced Mixer |
| MCA1-60+ | 7 | 1600 | 6000 | 1600 | 6000 | DC | 2000 | 6.3 | 32 | 17 | 1 | 9 | Double Balanced Mixer |
| MCA1-60LH+ | 10 | 1700 | 6000 | 1700 | 6000 | DC | 2000 | 6.4 | 35 | 21 | 5 | 13 | Double Balanced Mixer |

**Table 5.** Specifications of the ADE-MH35+ mixer

| Model Number | LO Level (dBm) | RF Freq. Low (MHz) | RF Freq. High (MHz) | LO Freq. Low (MHz) | LO Freq. High (MHz) | IF Freq. Low (MHz) | IF Freq. High (MHz) | Conversion Loss (dB) Typ. | LO-RF Isolation (dB) Typ. | LO-IF Isolation (dB) Typ. | Input P1dB (dBm) Typ. | Input IP3 (dBm) Typ. | Subcategory |
|---|---|---|---|---|---|---|---|---|---|---|---|---|---|
| ADE-35MH | 13 | 5 | 3500 | 5 | 3500 | 5 | 2500 | 6.9 | 33 | 28 | 9 | 18 | Double Balanced Mixer |
| ADE-35MH+ | 13 | 5 | 3500 | 5 | 3500 | 5 | 2500 | 6.9 | 33 | 28 | 9 | 18 | Double Balanced Mixer |
| ADE-42MH+ | 13 | 5 | 4200 | 5 | 4200 | 5 | 3500 | 7.5 | 29 | 26 | 9 | 17 | Double Balanced Mixer |

## 3.2 Amplifiers

TSS-53LNB+ from Mini-Circuits is selected for the LNA stage. This component has a noise figure of about 1.5 dB and a gain of about 21 dB in the frequency range of 2.7 GHz - 3.1 GHz. The low noise figure and high linearity (as implied by the high IIP3) make the TSS-53LNB+ an attractive option for the low-noise amplifier stage in the receiver RF chain, as shown in Table 6 [12].

**Table 6.** Specifications of the TSS-53LNB+ as LNA

| Model Number | F Low (MHz) | F High (MHz) | Gain (dB) Typ. | NF (dB) Typ. | P1dB(dBm) Typ. | OIP3 (dBm) Typ. | Input VSWR (:1) Typ. | Output VSWR (:1) Typ. | Voltage (V) | Current (mA) |
|---|---|---|---|---|---|---|---|---|---|---|
| TSS-23HLN+ | 30 | 2000 | 21.8 | 1.4 | 28.5 | 42.6 | 1.92 | 1.67 | 8 | 236 |
| TSS-23HLN-D+ | 30 | 2000 | 21.8 | 1.4 | 28.5 | 42.6 | 1.92 | 1.67 | 8/5/3 | 236/139/74 |
| TSS-23LN+ | 30 | 2000 | 21.5 | 1.2 | 24.1 | 36.4 | 1.92 | 1.67 | 5/3 | 139/74 |
| TSS-53LNB+ | 500 | 5000 | 21.7 | 1.4 | 20.6 | 33.9 | 1.46 | 1.33 | 5 | 82 |
| TSS-53LNB-D+ | 500 | 6000 | 21.4 | 1.3 | 19.4 | 35 | 1.2 | 1.4 | 5 | 82 |
| TSS-53LNB3+ | 500 | 5000 | 18.4 | 1.5 | 14.9 | 25 | 1.63 | 1.26 | 3 | 42 |
| TSY-13LNB+ | 30 | 1000 | 14.7 | 1.2 | 17.1 | 26.4 | 1.5 | 1.3 | 2.7 | 7.7 |
| TSY-172LNB+ | 30 | 1700 | 13.1 | 1.4 | 17.5 | 24.7 | 1.9 | 1.5 | 2.7 | 7.7 |

In an RF chain of a receiver, the selection of components requires a careful balance between noise and linearity performance. At the foremost stages, the noise factor is the primary concern, while towards the end of the chain, parameters related to the linearity of the chain, such as IIP3 and P1dB, become more critical [7]. When selecting the IF1 amplifier, it is significant to ensure that the chosen component performs well within the frequency range of 1785 MHz to 1815 MHz. As the RF chain progresses, the focus should shift towards prioritizing the linearity of the selected parts over their noise figure [7] [8]. According to Figure 13, the IF1 amplifier stage is after the first mixer. For this stage, the GALI-74+ amplifier from Mini-Circuits could be an appropriate choice [13]. While this amplifier has a higher noise figure than the previous TSS-53LNB+ amplifier in the chain, it offers superior linearity performance with a higher OIP3, as depicted in Table 7. Due to its relatively high noise figure of 2.7 dB, the GALI-74+ cannot be considered a low-noise amplifier. However, its impact on the overall noise figure of the receiver's RF chain is trivial due to the Friis formula [7] [8].

**Table 7.** Specifications of the GALI-74+ amplifier

| Model Number | F Low (MHz) | F High (MHz) | Gain (dB) Typ. | NF (dB) Typ. | P1dB(dBm) Typ. | OIP3 (dBm) Typ. | Input VSWR (:1) Typ. | Output VSWR (:1) Typ. | Voltage (V) | Current (mA) |
|---|---|---|---|---|---|---|---|---|---|---|
| GALI-55+ | DC | 4000 | 18.5 | 3.3 | 15.5 | 28.5 | 1.25 | 1.3 | 4.3 | 50 |
| GALI-59+ | DC | 5000 | 18.3 | 4.3 | 17.6 | 33.3 | 1.6 | 1.5 | 4.8 | 65 |
| GALI-74+ | DC | 1000 | 21.8 | 2.7 | 18.3 | 38 | 1.2 | 1.6 | 4.8 | 80 |
| GALI-84+ | DC | 6000 | 16.7 | 4.4 | 21 | 37.4 | 1.4 | 2.1 | 5.8 | 100 |
| GALI-S66+ | DC | 3000 | 18.2 | 2.4 | 3.3 | 19.1 | 1.1 | 1.2 | 3.5 | 16 |

When selecting the IF2 amplifier, as mentioned earlier, linearity considerations become more significant than the previous two amplifiers in the chain. As the receiver components get closer to the final stage, their nonlinear effects have more influence on the overall linearity of the entire RF chain. Therefore, the AH201 amplifier from TriQuint WJ can be a suitable option, as shown in Table 8 [14]. This amplifier has an OIP3 of around 45 dBm and a P1dB of about 29 dBm in the desired 57.5-62.5 MHz frequency range, Table 9, which is significantly better in terms of linear performance than the previous two amplifiers [14].

Placing a Variable Gain Amplifier (VGA) at the end of the RF chain in Figure 13 can be lucrative, providing more design flexibility. Considering the 57.5-62.5 MHz operating frequency range, the ADL5330 VGA appears to be a suitable and accessible option, Table 10 [15]. The gain of the amplifier can be varied from -35 dB to 22 dB by applying a control voltage between 0.1 V and 1.4 V at 100 MHz. The amplifier's OIP3 and P1dB values of around 38 dBm and 22 dBm, respectively, are acceptable for the final stage of the RF receiver [15].

**Table 8.** Specifications of the AH201 amplifier

**Specifications**

| Parameters | Units | Min | Typ | Max |
|---|---|---|---|---|
| Frequency Range | MHz | 50 | 900 | 2200 |
| Gain | dB | | 17 | |
| Input Return Loss | dB | | 20 | |
| Output Return Loss | dB | | 18 | |
| Output P1dB | dBm | +29 | +30 | |
| Output IP3 | dBm | +45 | +47 | |
| Noise Figure | dB | | 2.5 | |
| IS-95 Channel Power[3] @ -45dBc ACPR | dBm | | +24 | |
| Operating Current Range | mA | 310 | 350 | 390 |
| Supply Voltage | V | | +11 | |
| Thermal Resistance | °C / W | | | 17.5 |
| Junction Temperature[4] | °C | | | 160 |

Test conditions unless otherwise noted.
1. T = 25ºC, Vdd = 11 V, Frequency = 900 MHz in an application circuit.
2. 3OIP measured with two tones at an output power of +10 dBm/tone separated by 10 MHz. The suppression on the largest IM3 product is used to calculate the 3OIP using a 2:1 rule.
3. IS-95, 9 Channels Forward, Pk/Avg Ratio = 11.5 dB at a .001% probability 750 kHz offset, 30 kHz bandwidth, Channel BW = 1.23 MHz, frequency = 860 MHz
4. The maximum junction temperature ensures a minimum MTTF rating of 1 million hours of usage. The MTTF plot and more information is located on the WJ App Note "Temperature Effects on Reliability for the AH201" located on the WJ website.

**Typical Performance**

| Parameters | Units | Typical | | |
|---|---|---|---|---|
| Frequency | MHz | 900 | 1900 | 2140 |
| Gain | dB | 17 | 15 | 15 |
| Input Return Loss | dB | 20 | 9.1 | 9.2 |
| Output Return Loss | dB | 18 | 15 | 18 |
| Output P1dB | dBm | +30.0 | +29.7 | +29.4 |
| Output IP3 | dBm | +47 | +46 | +46 |
| Noise Figure | dB | 2.5 | 3.8 | 4.2 |
| IS-95 Channel Power[3] @ -45dBc ACPR | dBm | +24 | +24 | +24 |
| Supply Bias | | +11 V @ 350 mA | | |

Typical parameters reflect performance in an application circuit.

**Table 9.** S-parameters of the AH201 in different frequencies

S-Parameters ($V_{DS}$ = +10V, $I_{DS}$ = 350 mA, T = 25°C, unmatched device in a 50 Ω system)

| Freq (MHz) | S11 (dB) | S11 (ang) | S21 (dB) | S21 (ang) | S12 (dB) | S12 (ang) | S22 (dB) | S22 (ang) |
|---|---|---|---|---|---|---|---|---|
| 200 | -18.13 | -141.77 | 17.77 | 156.16 | -21.60 | -12.80 | -15.22 | 158.80 |
| 400 | -14.01 | -151.43 | 17.47 | 134.85 | -22.01 | -27.05 | -15.91 | 138.77 |
| 600 | -11.32 | -161.69 | 17.03 | 113.80 | -22.50 | -40.65 | -19.50 | 119.22 |
| 800 | -9.60 | -175.43 | 16.52 | 93.61 | -23.08 | -54.53 | -28.09 | 86.76 |
| 1000 | -8.38 | 170.85 | 15.98 | 74.11 | -23.75 | -66.32 | -28.42 | -67.31 |
| 1200 | -7.72 | 156.06 | 15.50 | 55.43 | -24.58 | -80.56 | -18.92 | -96.94 |
| 1400 | -7.51 | 140.70 | 15.09 | 37.11 | -25.23 | -94.35 | -14.49 | -115.52 |
| 1600 | -7.69 | 123.37 | 14.78 | 18.41 | -25.58 | -107.34 | -11.68 | -131.45 |
| 1800 | -8.42 | 102.05 | 14.63 | -0.87 | -26.55 | -124.66 | -9.76 | -146.54 |
| 2000 | -9.90 | 76.87 | 14.53 | -21.84 | -26.58 | -144.30 | -7.87 | -160.86 |
| 2200 | -12.60 | 27.48 | 14.50 | -45.98 | -26.08 | -170.04 | -7.03 | -176.78 |
| 2400 | -11.30 | -52.35 | 14.07 | -74.59 | -25.82 | 158.59 | -7.07 | 169.02 |
| 2600 | -6.47 | -110.35 | 12.82 | -105.64 | -25.42 | 125.63 | -7.82 | 159.81 |
| 2800 | -3.46 | -149.16 | 10.54 | -135.77 | -25.19 | 97.73 | -8.39 | 158.80 |
| 3000 | -2.00 | -178.37 | 7.60 | -161.74 | -25.67 | 70.96 | -7.81 | 158.95 |

**Table 10.** Specifications of the ADL5330 amplifier as VGA

| Parameter | Conditions | Min | Typ | Max | Unit |
|---|---|---|---|---|---|
| GENERAL | | | | | |
| Usable Frequency Range | | 0.01 | | 3 | GHz |
| Nominal Input Impedance | Via 1:1 single-sided-to-differential balun | | 50 | | Ω |
| Nominal Output Impedance | Via 1:1 differential-to-single-sided balun | | 50 | | Ω |
| 100 MHz | | | | | |
| Gain Control Span | ±3 dB gain law conformance | | 58 | | dB |
| Maximum Gain | $V_{GAIN}$ = 1.4 V | | 23 | | dB |
| Minimum Gain | $V_{GAIN}$ = 0.1 V | | −35 | | dB |
| Gain Flatness vs. Frequency | ±30 MHz around center frequency, $V_{GAIN}$ = 1.0 V (differential output) | | 0.09 | | dB |
| Gain Control Slope | | | 20.7 | | mV/dB |
| Gain Control Intercept | Gain = 0 dB, gain = slope ($V_{GAIN}$ − intercept) | | 0.88 | | V |
| Input Compression Point | $V_{GAIN}$ = 1.2 V | | 1.8 | | dBm |
| Input Compression Point | $V_{GAIN}$ = 1.4 V | | −0.3 | | dBm |
| Output Third-Order Intercept (OIP3) | $V_{GAIN}$ = 1.4 V | | 38 | | dBm |
| Output Noise Floor[1] | 20 MHz carrier offset, $V_{GAIN}$ = 1.4 V | | −140 | | dBm/Hz |
| Noise Figure | $V_{GAIN}$ = 1.4 V | | 7.8 | | dB |
| Input Return Loss[2] | 1 V < $V_{GAIN}$ < 1.4 V | | −12.8 | | dB |
| Output Return Loss[2] | | | −15.5 | | dB |

### 3.3 Filters

The general diagram of the RF chain in Figure 13 indicates that after the first amplifier (TSS-53LNB+), there is a passband filter. This filter, with the part number BFCN-2900+, is from the Mini-Circuits Company, Table 11 [16]. It is a ceramic passband filter that serves as a pre-selector with a bandwidth of 2.7 GHz to 3.1 GHz. Also, the filter offers a compact size and good thermal stability.

The subsequent IF filter in the RF chain was designed with a center frequency of 1.8 GHz and a bandwidth of 30 MHz based on the analysis of spurious signal considerations discussed in section 2. However, finding a commercially available filter with these precise specifications proved challenging and costly.

**Table 11.** Specifications of the BFCN-2900+ filter

| Model Number | Passband F1 (MHz) | Passband F2 (MHz) | Stopband F3 (MHz) | Rejection @ F3 (dB) | Stopband F4 (MHz) | Rejection @ F4 (dB) | Filter Type |
|---|---|---|---|---|---|---|---|
| BFCN-2850+ ez | 2750 | 2950 | 1500 | 20 | 4300 | 20 | Band Pass |
| BFCN-2900+ ez | 2700 | 3100 | 1850 | 20 | 4200 | 20 | Band Pass |
| BFCN-2910+ ez | 2850 | 2970 | 1600 | 20 | 4200 | 20 | Band Pass |
| BFCN-2975+ ez | 2570 | 3440 | DC-1700 | 20 | 4000-7500 | 20 | Band Pass |

Thus, we decided to design the filter exclusively. One of the appropriate types of microwave filters that is relatively straightforward to design for such narrow-band applications, with a relative bandwidth of less than 20%, is the parallel-coupled line or coupled-line filter. By properly designing the coupled-line structure to exploit the even-mode and odd-mode characteristics, filter designers can achieve the desired frequency response and impedance matching for the coupled-line filter. However, the design method inherently carries an error of up to 5%, which is a significant factor given the stringent requirements for this filter [17]. The optimal relative bandwidth for this application is around 1.6% ($(\frac{30MHz}{1.8GHz}) = \%1.6$).

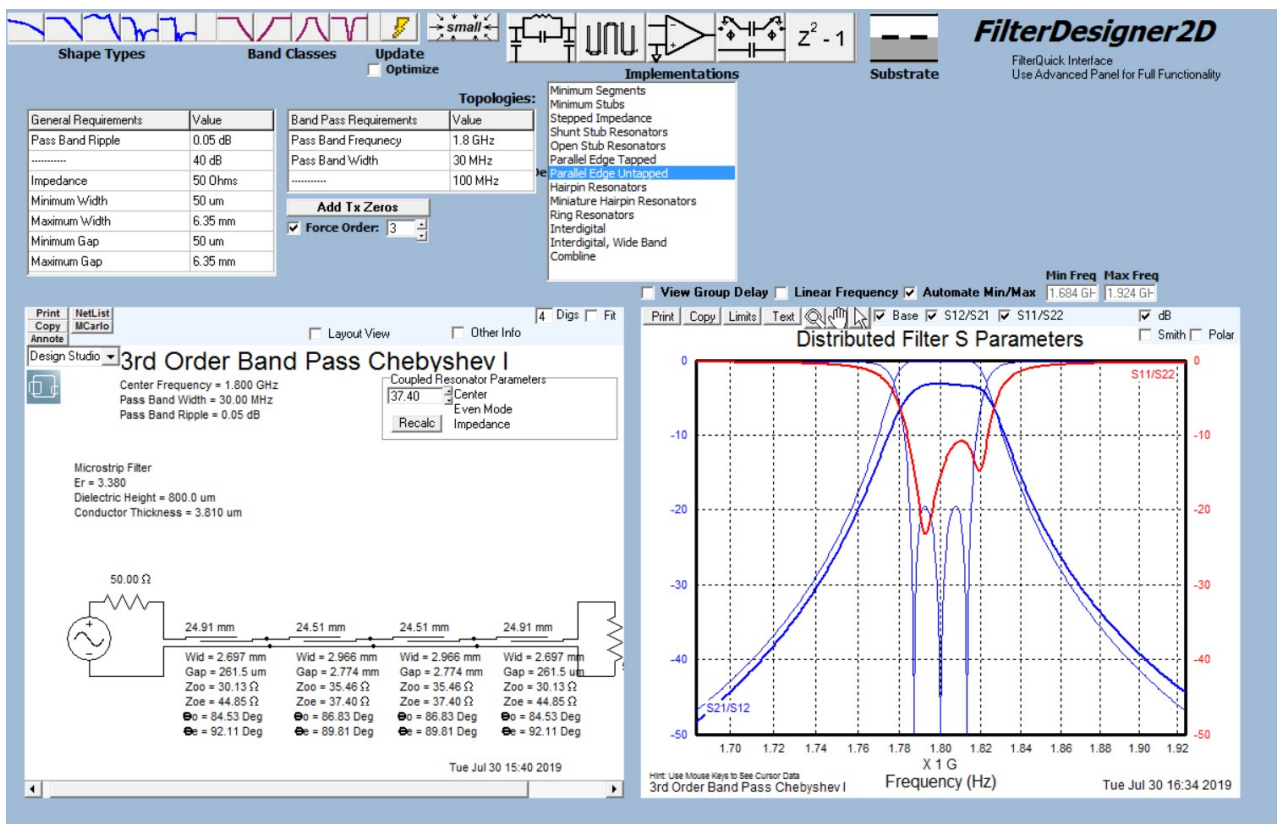

**Fig. 14.** Using the filter design tool of CST Studio Suite, the design and analysis tasks are performed simultaneously.

One of the best approaches to a more accurate filter design is to utilize full-wave analysis methods. The full-wave methods are generally the most precise techniques for high-frequency circuit analysis, although with augmented complexity. One such full-wave analysis method is the Finite-Difference Time-Domain (FDTD) technique employed by the FilterDesigner2D module as a tool within the CST Studio Suite. Figure 14 presents the output of this GUI-based design tool. It is worth noticing that the substrate material used in this design is RG4003, attributed to a 30-mil thickness.

This GUI allows both implementation and real-time analysis of filters based on user-specified parameters such as filter type, order, bandwidth, and center frequency. The filter topology selected for the given operating frequency range is coupled-line, and a distributed implementation approach is considered. To implement the 30 MHz bandwidth filter with a 1.8 GHz center frequency, a 3rd-order Chebyshev filter type is used.

Figure 14 plots the S-parameters of this filter design. The pale blue lines represent the $S_{11}$, $S_{22}$, $S_{12}$, and $S_{21}$ responses of the ideal Chebyshev filter. The solid blue line shows the analyzed $S_{21}$ and $S_{12}$ values after the implementation using transmission lines.

Similarly, the solid red line indicates the analyzed $S_{11}$ and $S_{22}$ values for the coupled-line filter design. As evident from the solid lines in the plot, the designed filter exhibits an acceptable frequency response within the desired operating frequency range. Also, Figure 14 presents the output of the filter design process using the GUI, which includes the length, spacing, and other characteristics of the coupled transmission lines. Given the derived filter dimensions and meeting the required frequency response, the coupled-lined filter seems feasible for implementation. However, it is pertinent to confirm the reliability of this full-wave design method. To this end, it was decided to re-evaluate the filter parameters designed using the CST FilterDesigner2D tool, this time employing a different full-wave analysis method. In this regard, the Layout section of the ADS software was leveraged, as it utilizes the Method of Moments (MoM) full-wave technique to analyze microwave circuits. Figures 15 and 16 show the filter designed from the CST FilterDesigner2D, simulated in the ADS software's Schematic and Layout, respectively. As depicted in Figure 15, the MSub parameters were configured to match the Rogers4003 substrate with a 30-mil thickness, consistent with the previous CST FilterDesigner2D analysis. Figure 17 presents the output of the MoM analysis conducted on this filter design.

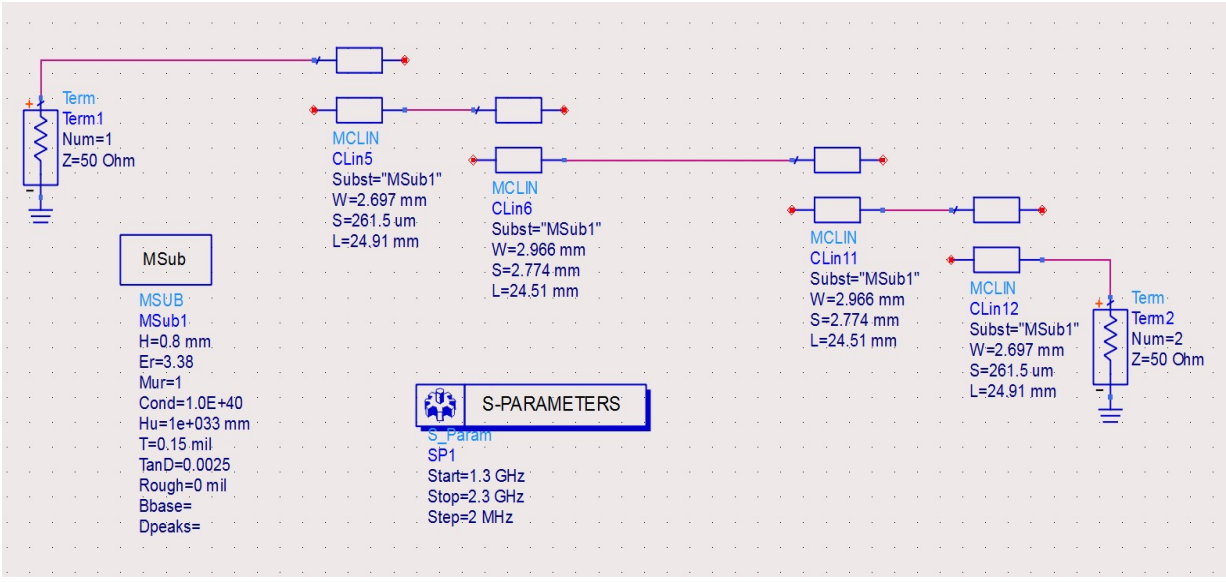

**Fig 15.** Distributed coupled-line filter designed using the CST FilterDesigner2D tool depicted in the ADS Schematic.

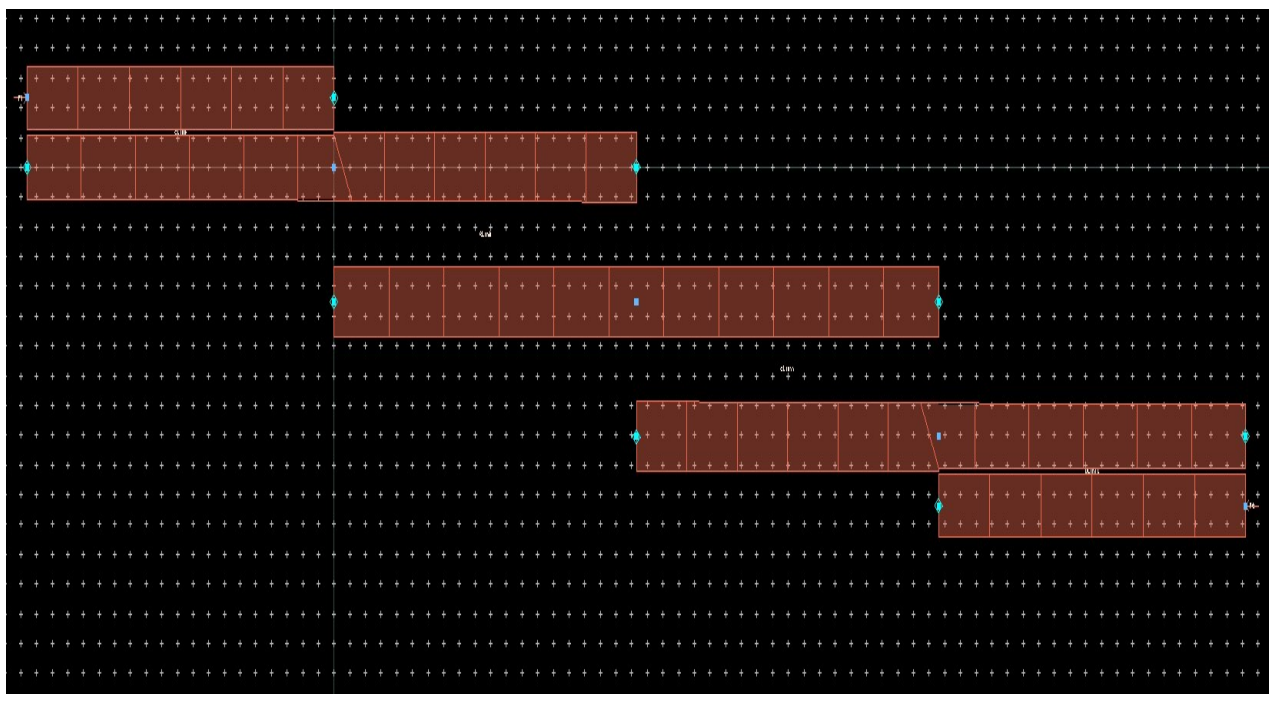

**Fig 16.** Distributed coupled-line filter designed using the CST FilterDesigner2D tool presented in the ADS Layout.

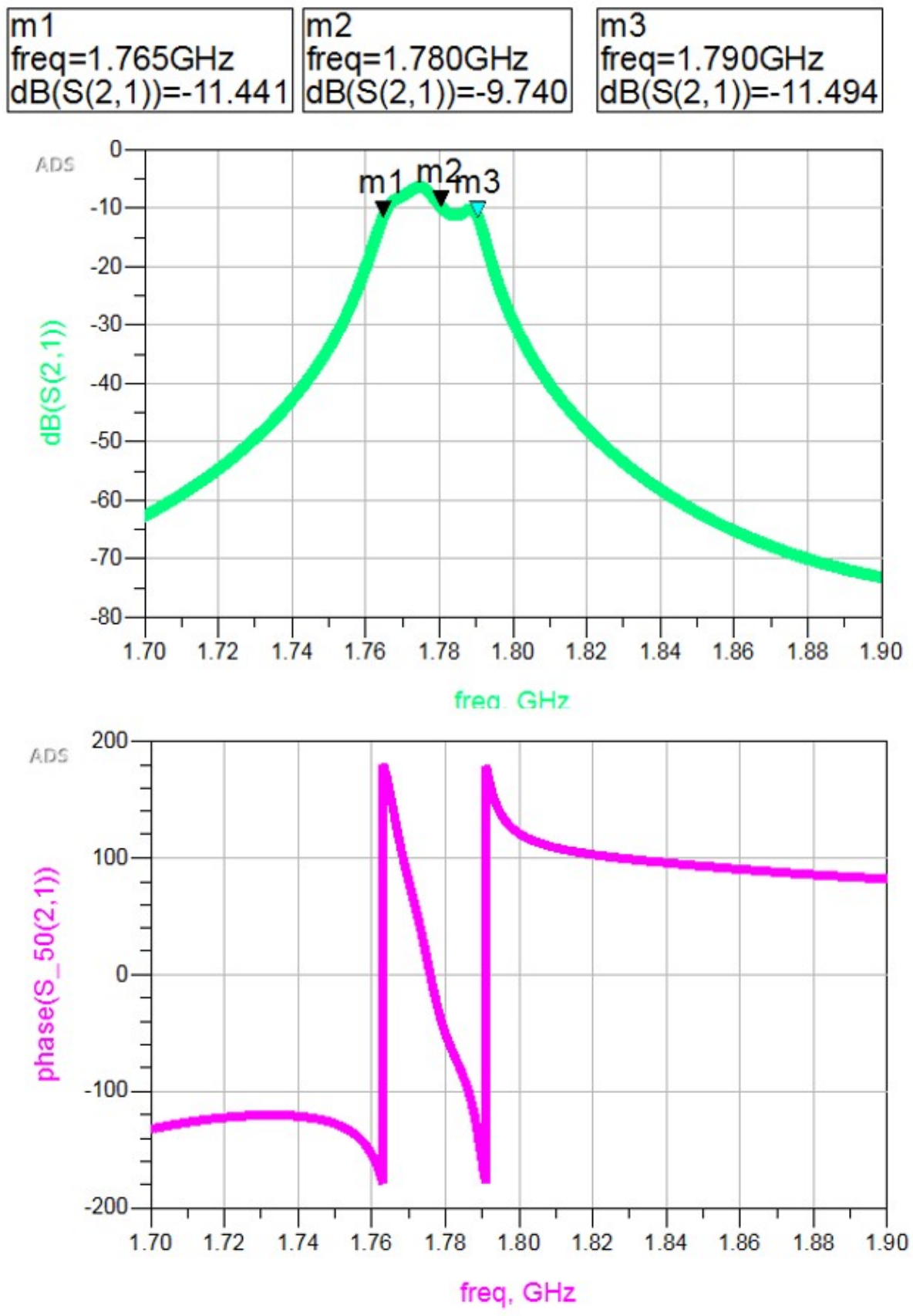

**Fig 17.** Phase response and magnitude of the MoM in ADS Layout.

As demonstrated in Figure 17, both the phase response and magnitude response reveal a linear trend around 1.78 GHz. Compared to the results presented in Figure 14, it is conspicuous that the output of the distributed filter designed using the CST FilterDesigner2D tool in Figure 14 differs by approximately 20 MHz from the result obtained through the Layout analysis in the ADS in Figure 17. Knowing the filter's bandwidth is 30 MHz, such a discrepancy cannot be overlooked. Thus, a comprehensive reversal of the design and verification process is required to compare these two analysis approaches more accurately. To this end, in the first step, the scheme entails designing the filter dimensions so that the center frequency of the linear part of the frequency (magnitude and phase) response in the ADS Layout output aligns with 1.8 GHz, which serves as the central frequency of the passband. Subsequently, the obtained design will be verified using the FDTD analysis in the CST Studio Suite.

First, a preliminary design was considered using the analytical methods of microwave filter design, which provided the even and odd mode impedance values [18]. Next, the desired microstrip filter dimensions were obtained using the LineCalc module within the ADS. Finally, the transmission line dimensions were then refined by employing the tuning/optimization feature in ADS, ensuring that the phase and amplitude responses from the MoM analysis in the ADS Layout section matched the desired specifications - the center frequency of 1.8 GHz and a bandwidth of around 30 MHz.

To initiate the design process, let's assume the goal is to design a 3rd-order Chebyshev filter with a 10 MHz bandwidth and a 1.8 GHz center frequency. The coefficient values in (1) were obtained by referencing the available design tables for this 3rd-order Chebyshev filter design [18]:

$$g_0 = g_4 = 1,\ g_1 = g_3 = 1.5963\ ,\ g_2 = 1.0967. \tag{1}$$

The impedance values for the desired odd and even were determined by obtaining the coefficients using (2), (3), and (4). It is worth noting that the initial assumption for this design was a filter ripple value of 0.5 dB. Consequently, the following results were obtained [18]:

$$Z_0 J_1 = \sqrt{\frac{\pi \times \Delta\left(= \dfrac{BW}{f_{CF}}\right)}{2g_1}} = \sqrt{\frac{\pi \times \Delta\left(= \dfrac{BW}{f_{CF}}\right)}{2g_3}} = \sqrt{\frac{\pi \times \dfrac{10MHz}{1.8GHz}}{2 \times 1.5963}} = 0.0739 = Z_0 J_4 \tag{2}$$

$$Z_0 J_2 = \frac{\pi \times \Delta}{2\sqrt{g_1 g_2}} = \frac{\pi \times \Delta}{2\sqrt{g_2 g_3}} = Z_0 J_3 = 0.0066 \tag{3}$$

$$\begin{cases} Z_{0e_n} = Z_0\left[1 + J_n Z_0 + \left(J_n Z_0\right)^2\right] \\ Z_{0o_n} = Z_0\left[1 - J_n Z_0 + \left(J_n Z_0\right)^2\right] \end{cases} \xrightarrow{Z_0 = 50\Omega} \begin{cases} Z_{0e_1} = Z_{0e_4} = 53.97\Omega \\ Z_{0o_1} = Z_{0o_4} = 46.58\Omega \\ Z_{0e_2} = Z_{0e_3} = 50.33\Omega \\ Z_{0o_2} = Z_{0o_3} = 49.67\Omega \end{cases} \tag{4}$$

The initial dimensions of the microstrip filter were obtained using the LineCalc module within the ADS and specifying a 30-mil Rogers4003 substrate.

Following the determination of the dimensions for the four distinct sections of this microstrip filter using the ADS tuning feature, the dimensions of the microstrip filter were adjusted to achieve a center frequency of 1.8 GHz and a bandwidth of approximately 30 MHz in the magnitude and phase response of the Layout section of the ADS, as previously mentioned. Figure 20 presents the schematic of such a distributed circuit after the tuning process in the ADS and the acquisition of the new microstrip line dimensions.

Once the dimensions of the microstrip filter were obtained, a MoM analysis was conducted on the filter in the ADS Layout. The representations of the designed filter are available within the ADS Schematic and Layout in Figures 20 and 21. After the MoM analysis, the output is presented as the magnitude and phase response in Figure 22.

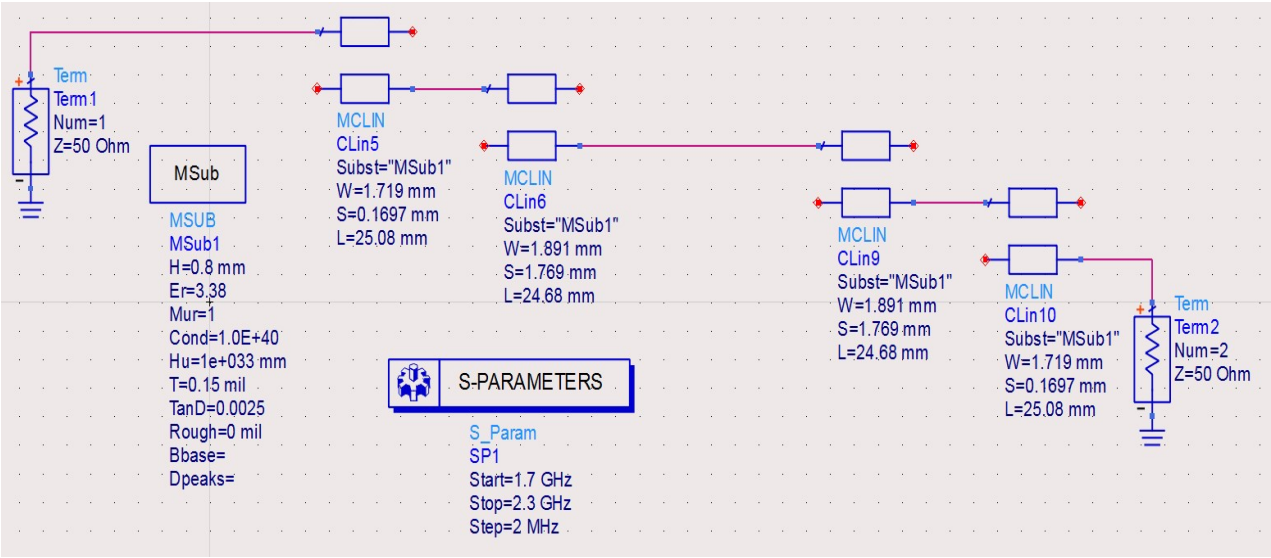

**Fig. 20.** Schematic of the filter designed following the tuning process in the ADS to deliver the specified response within the MoM analysis of the Layout section of the ADS.

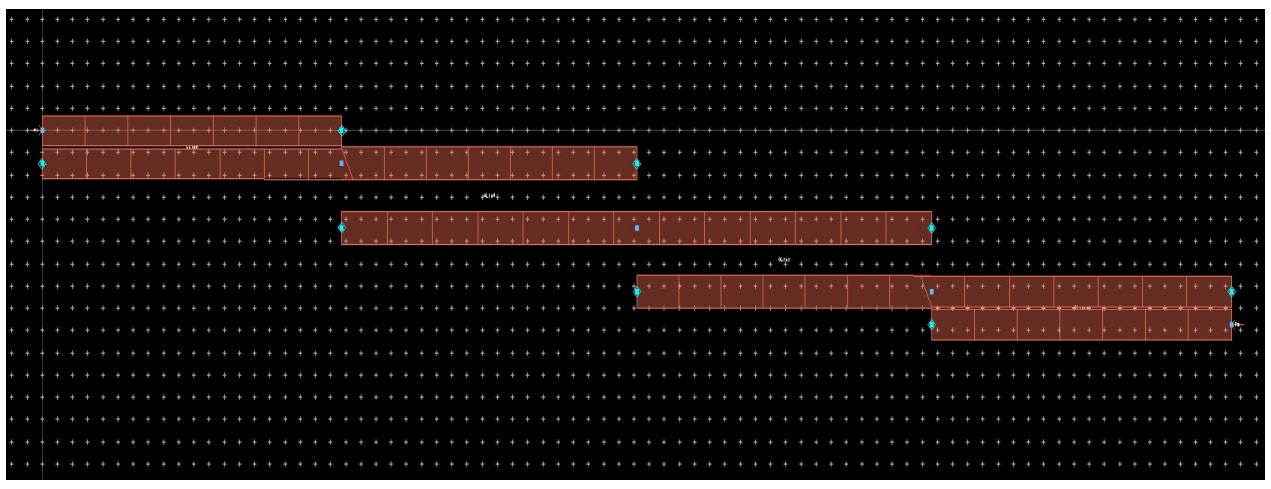

**Fig. 21.** Distributed coupled-line filter designed using even and odd mode analysis, refined through the tuning process in ADS, as shown in the Layout section.

As evident from Figure 22, the output of the MoM analysis performed on the Layout section indicates that the center frequency of the designed filter is about 1.8 GHz, with a bandwidth of 30 MHz. An evaluation for verification was conducted using the FilterDesigner2D section of the CST Studio on the newly coupled-line filter, designed using even and odd mode analysis and refined through the tuning process within the ADS. The newly coupled-line filter's dimensions from Figure 20 were used as input to the GUI in CST Studio, and the FDTD method was applied to analyze the filter. Figure 23 shows the resulting S-Parameters of such an analysis.

In Figure 23, the solid blue line represents the magnitude of the $S_{21}$ and $S_{12}$, and the solid red line shows the magnitude of the $S_{22}$ and $S_{11}$, obtained after analyzing the newly coupled-line filter using the FDTD method. Also, the pale blue lines in Figure 23 represent the $S_{11}$, $S_{22}$, $S_{12}$, and $S_{21}$ responses of the ideal Chebyshev filter. As evident from Figure 23, the central frequency of the designed filter is close to 1.82 GHz, which is shifted by approximately 20 MHz compared to the results shown in Figure 22. Comparing the results in Figure 22 and Figure 23 leads to the same conclusion drawn from the comparison of Figure 14 and Figure 17. Interestingly, if we formulate the design criteria using one of these methods and then evaluate the resulting design with the other approach, the analysis of the final design will differ by 20 MHz from the results obtained through the alternate method. Such a conclusion leads to the question of which of these two approaches is the more appropriate method to employ as a base for the design. Is the FDTD method in the FilterDesigner2D section of the CST Studio more accurate than the MoM method in the Layout section of the ADS?

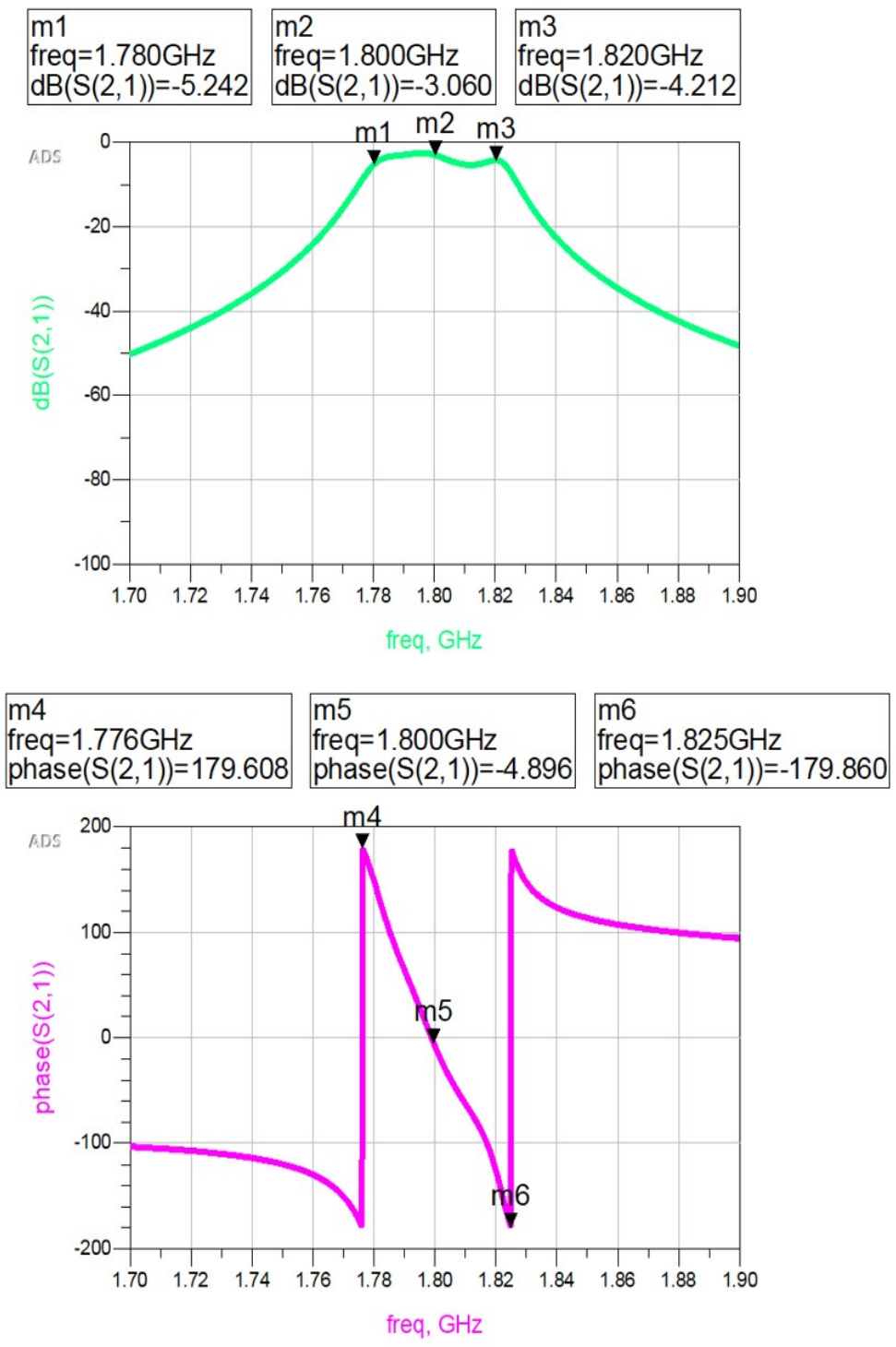

**Fig. 22.** Magnitude and phase response of the coupled-line filter, as shown in Figure 21, determined through MoM analysis in the Layout section of ADS.

Addressing this question requires carefully considering the practical conditions and the fabrication of printed circuits. High-frequency printed circuits often have a multi-layer structure, creating a heterogeneous medium. In such cases, the full-wave FDTD method is more appropriate for designing a more accurate output, as it can better accommodate the heterogeneous nature of the multi-layer structure. On the other hand, the MoM method is more suited for a homogeneous medium. Due to the above evaluation, the dimensions of the desired coupled-line filter should be based on the results obtained in Figure 16, which were derived using the FDTD method in the FilterDesigner2D section of the CST Studio, not the dimensions in Figure 20 or Figure 23.

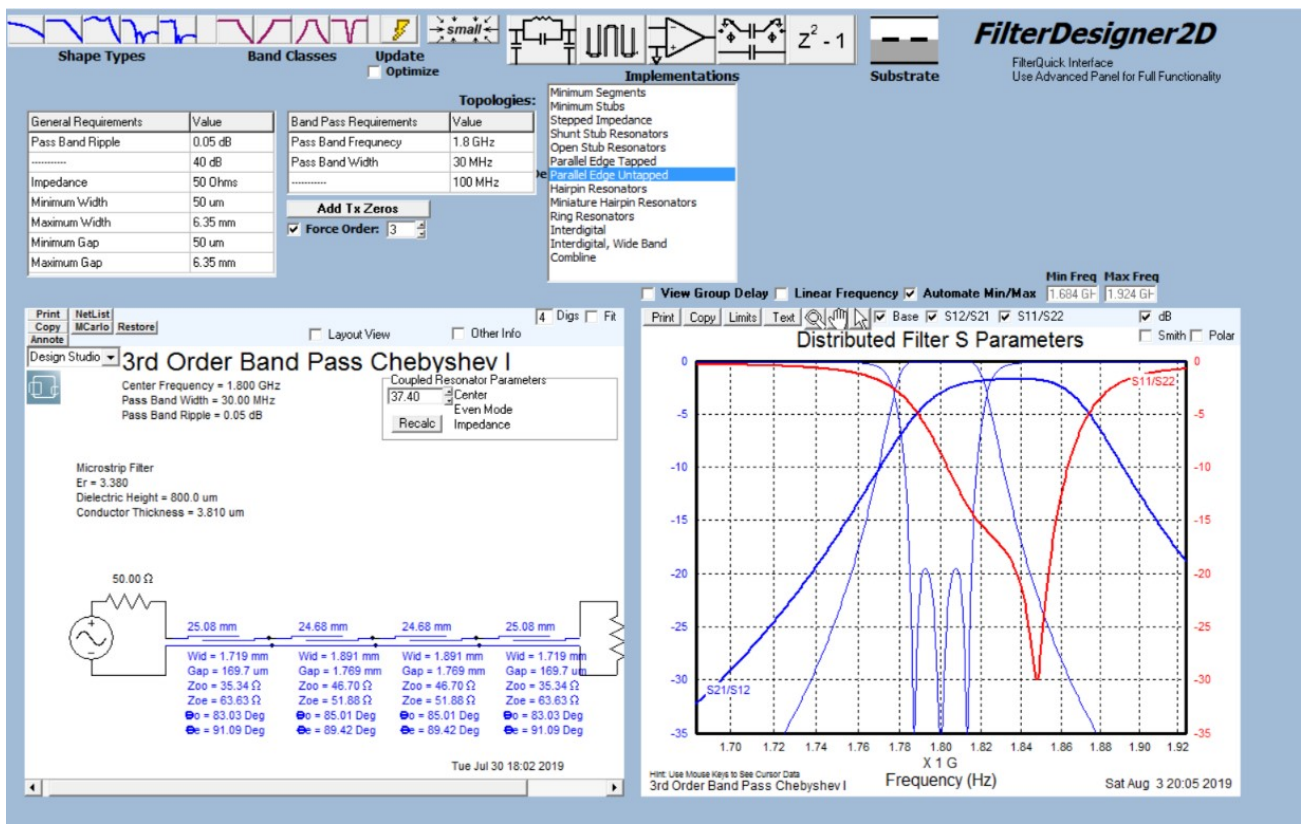

**Fig. 23.** Results of FDTD analysis on the newly designed coupled-line filter, utilizing even and odd mode analysis and optimized through tuning process in ADS software, obtained in the FilterDesigner2D section of CST Studio.

The following IF filter in the RF chain is the last one in the chain, and based on the frequency plan and Figure 13, the required bandwidth is approximately 5 MHz. Given the operating frequencies, the most cost-effective approach for the implementation is to use a filter designed with compact and passive components, specifically an LC filter. To this end, the DesignGuide section within the ADS could be accommodating. This section lets the user determine the filter type, order, and cutoff frequency and guides the implementation and design process. Figure 24 displays the graphical user interface associated with this DesignGuide functionality.

By opting for the most proximate standard practical inductor and capacitor values, the outcome is the component values derived during the circuit design process depicted in Figure 25, whose frequency response can be seen in Figure 26, respectively.

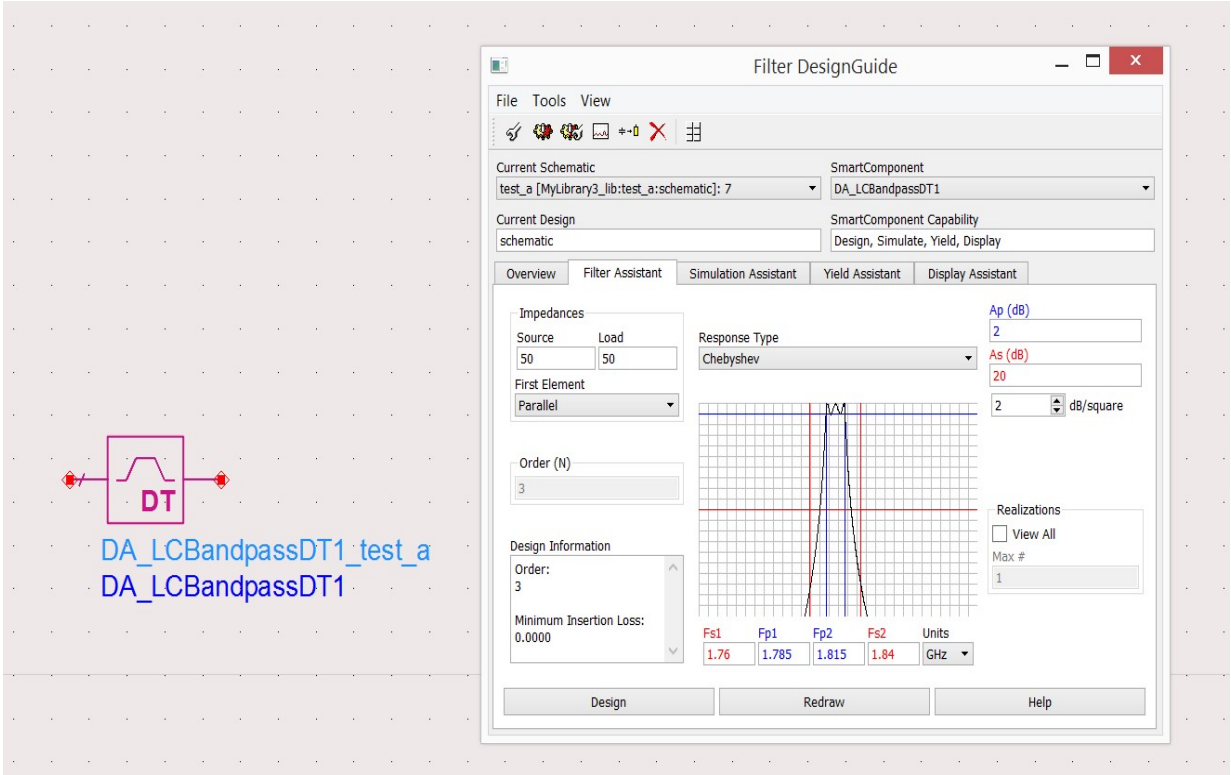

**Fig 24.** Designing lumped filters using the relevant tools available in the DesignGuide section of the ADS.

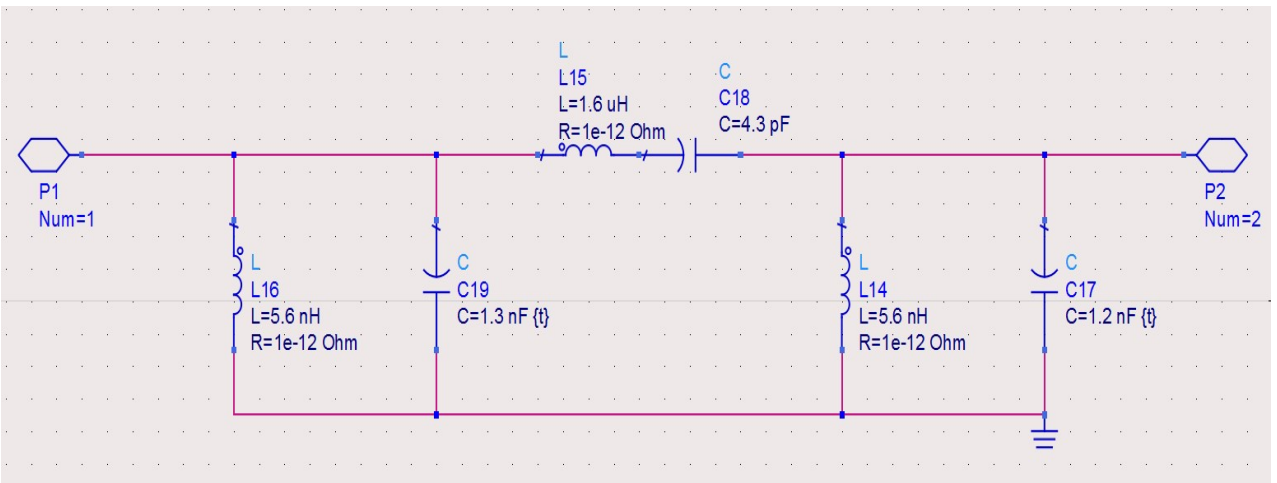

**Fig. 25.** Lumped filter completed using commercially available inductor and capacitor components with values close to design specifications.

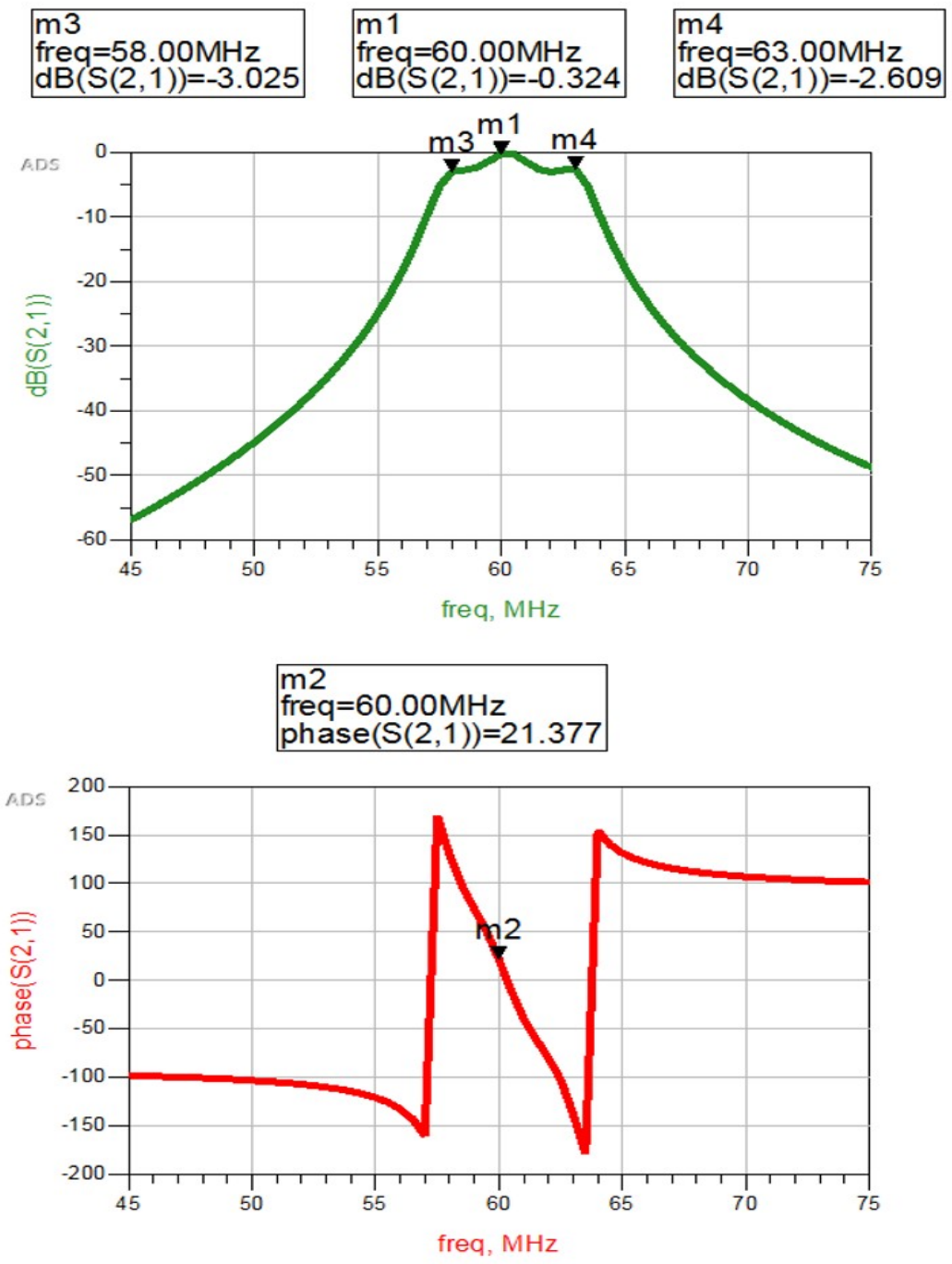

**Fig 26.** Frequency response of the schematic in Figure 25.

As shown in Figure 26, the phase and magnitude response of the output matches the desired response characteristics.

## 4. Comprehensive Analysis of The Designed Chain

Once the receiver components have been designed and selected, the complete RF chain is analyzed to verify if the desired specifications in Table 3 are met. The CASCADE software and ADS tools are utilized to evaluate the performance of the entire designed RF signal chain.

### 4.1 CASCADE Analysis

The software calculates the overall system parameters by sequentially cascading the individual components. Such a scheme enables isolating and observing the effect of the selected components on the various parameters of the complete signal chain.

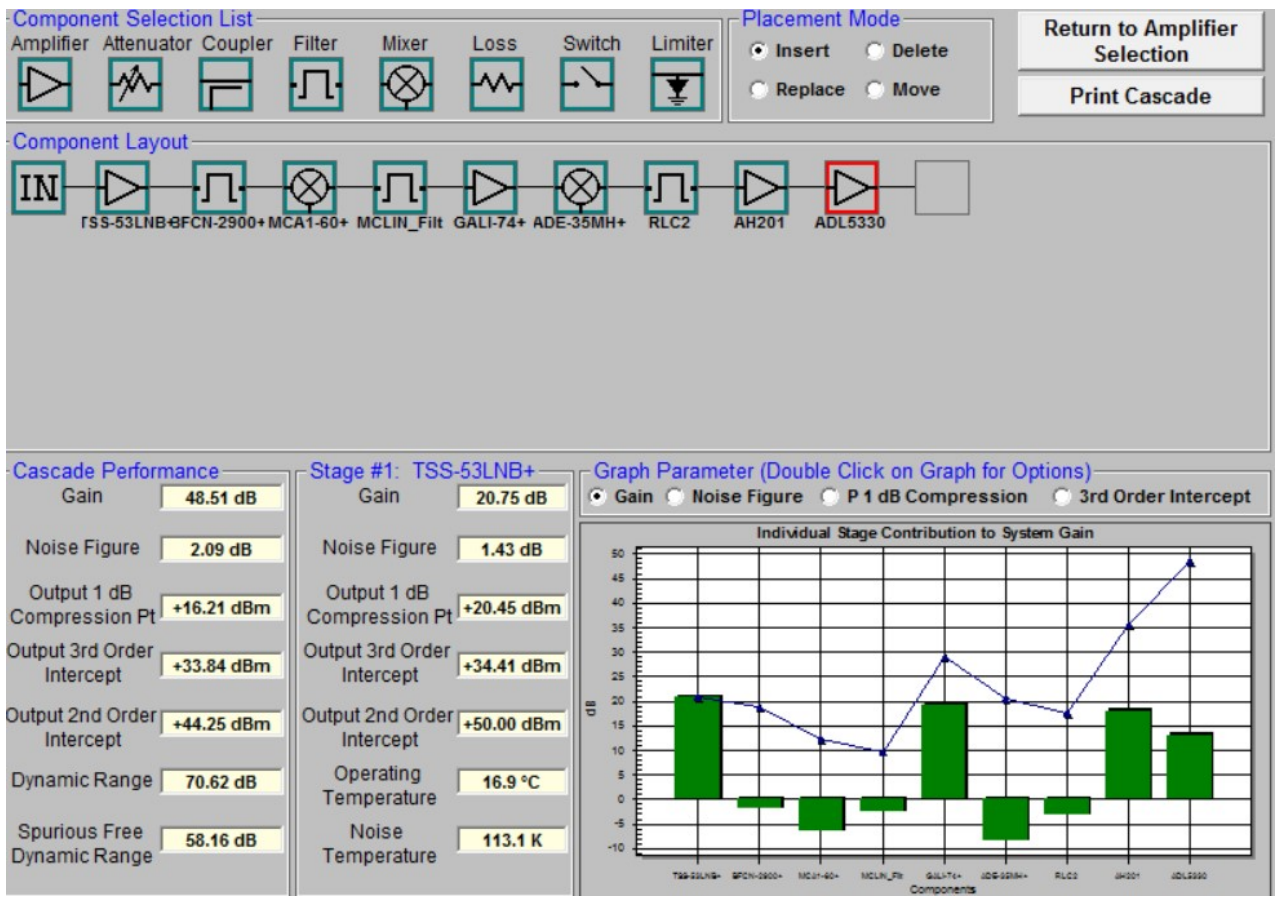

**Fig. 27.** Evaluating the designed RF chain using CASCADE software.

The analysis, Figure 27, shows that, excluding the gain and OIP3 values, the rest of the parameters, including the dynamic range, have met the desired specifications outlined in Table 1. The results demonstrate a pleasing match with the specifications. Also, the gain and OIP3 values are not significantly different from their desired values of 50 dB and 35 dBm, respectively.

Discrepancies exist between the results obtained through the CASCADE software and the design presented in sections 2 and 3. However, it is significant to note that although CASCADE offers a quick way to analyze a typical RF chain, its performance limitations can affect the reliability of its results. Primarily, the analysis is restricted to a single tone (frequency) without the ability to evaluate the performance of the chain's different components across different bandwidths. Also, the CASCADE software considers the RF chain's performance only with a single input signal at a determined power level. A thorough analysis of an RF chain requires simulation with different power levels to evaluate the system parameters accurately. Yet, the CASCADE software cannot adjust the input signal power level. Consequently, the results obtained from the CASCADE may not be reliable due to these limitations. Regarding such unreliability, careful consideration of these limitations is necessary when interpreting the data, and a more comprehensive analysis should be performed using alternative software.

### 4.2 Circuit Analysis

The ADS addresses two limitations of the CASCADE software by offering analysis across different frequency ranges and accommodating variable input conditions. One particularly appropriate analysis within ADS is the Link Budget analysis, developed for RF chain evaluation in transmitters. Such a tool enables the analysis of the entire RF chain or specific components as a circuit within the whole chain. Furthermore, individual components can be defined as blocks with their respective S-parameters, enabling analysis of these blocks in conjunction with the rest of the chain. Figure 28 depicts the schematic of the RF chain with the components selected in section 3, similar to Figure 13.

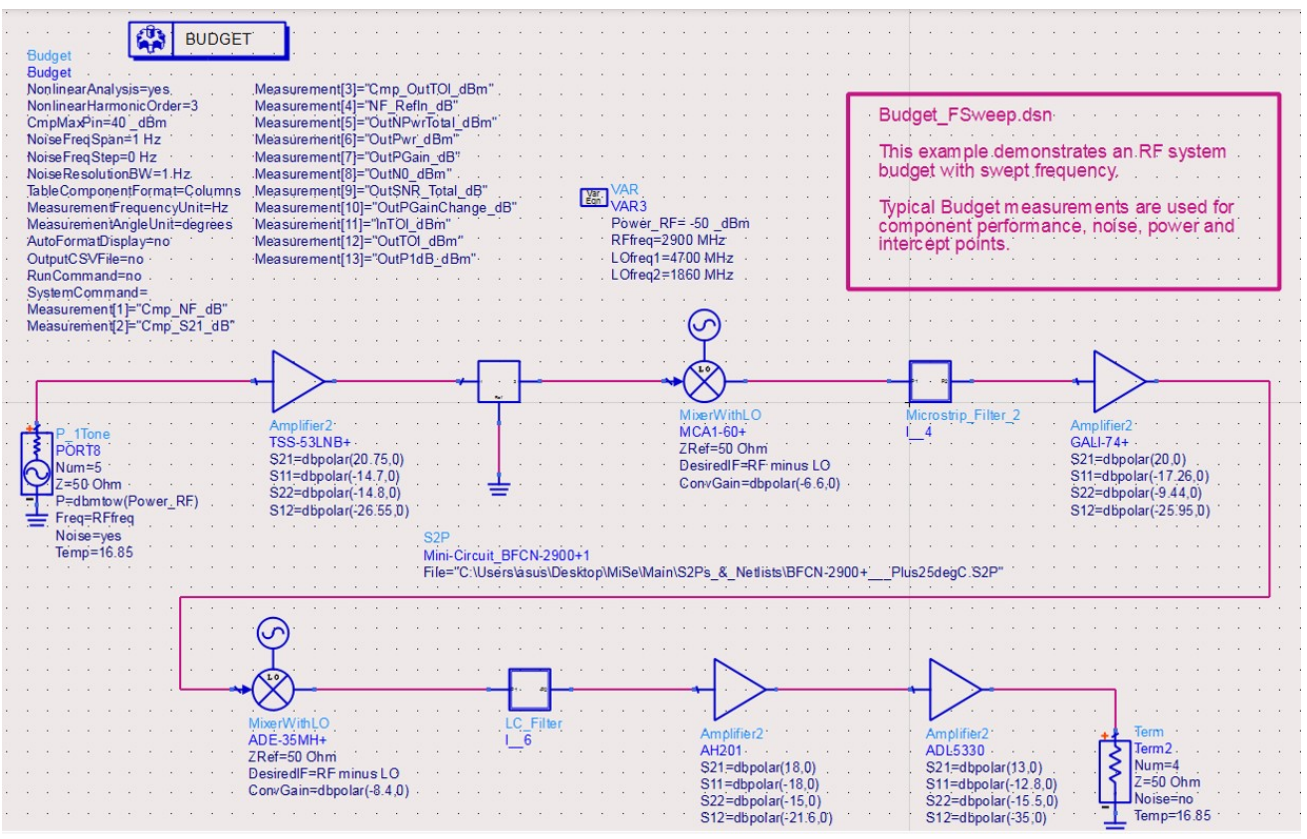

**Fig. 28.** Evaluating the designed RF chain using the ADS

In Figure 28, the amplifiers and mixers are modeled based on their datasheet specifications. The first filter, BFCN-2900+, utilizes an S2P file containing the S-parameter values from its datasheet for the operated frequency range. The first IF filter, block named I__4, includes the distributed microstrip filter circuit with paired lines, as shown in Figure 29. The frequency response of this filter in the desired range is depicted in Figure 30. The second IF filter, a block named I__6, contains the compact circuit design, as shown in Figure 31, the same as the filter in Figure 25.

The dimensions of the microstrip filter branches in the schematic shown in Figure 29 differ slightly from those in Figure 14 (or Figure 15). Such an alteration arises because the Link Budget section of the ADS utilizes S-parameter linear analysis for filters, and using the dimensions from Figure 14 leads to inaccurate results. Since the Link Budget section cannot perform FDTD analysis and is intended solely for assessing the overall performance of the whole blocks of the RF chain, the microstrip filter dimensions should align with the dimensions presented in Figure 29 (which rely on S-parameter linear analysis). However, for practical implementation of the same microstrip filter as a block in the RF chain, it is supposed that the dimensions from Figure 14 will render better results when measuring the overall specifications of this RF chain. If the Link Budget section could perform FDTD analysis, it would not require modifying the dimensions. Figure 30 displays the S-parameter analysis of the filter shown in Figure 29.

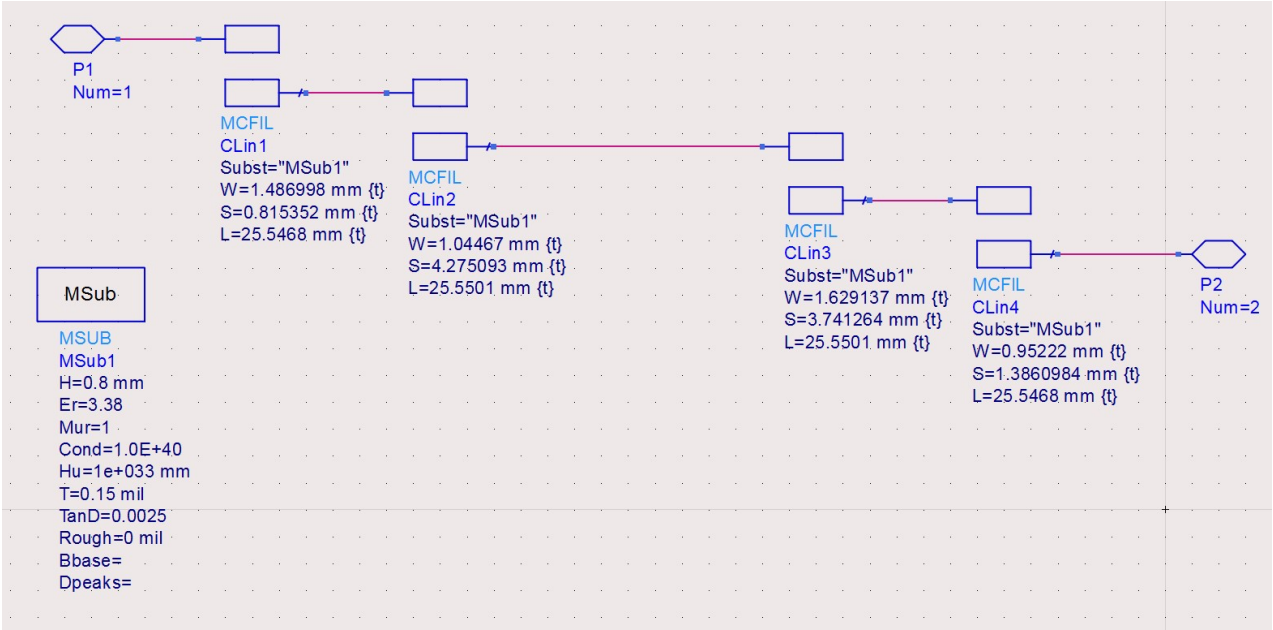

**Fig 29.** Microstrip filter placed through block titled I__4 in the RF chain of Figure 28 within ADS.

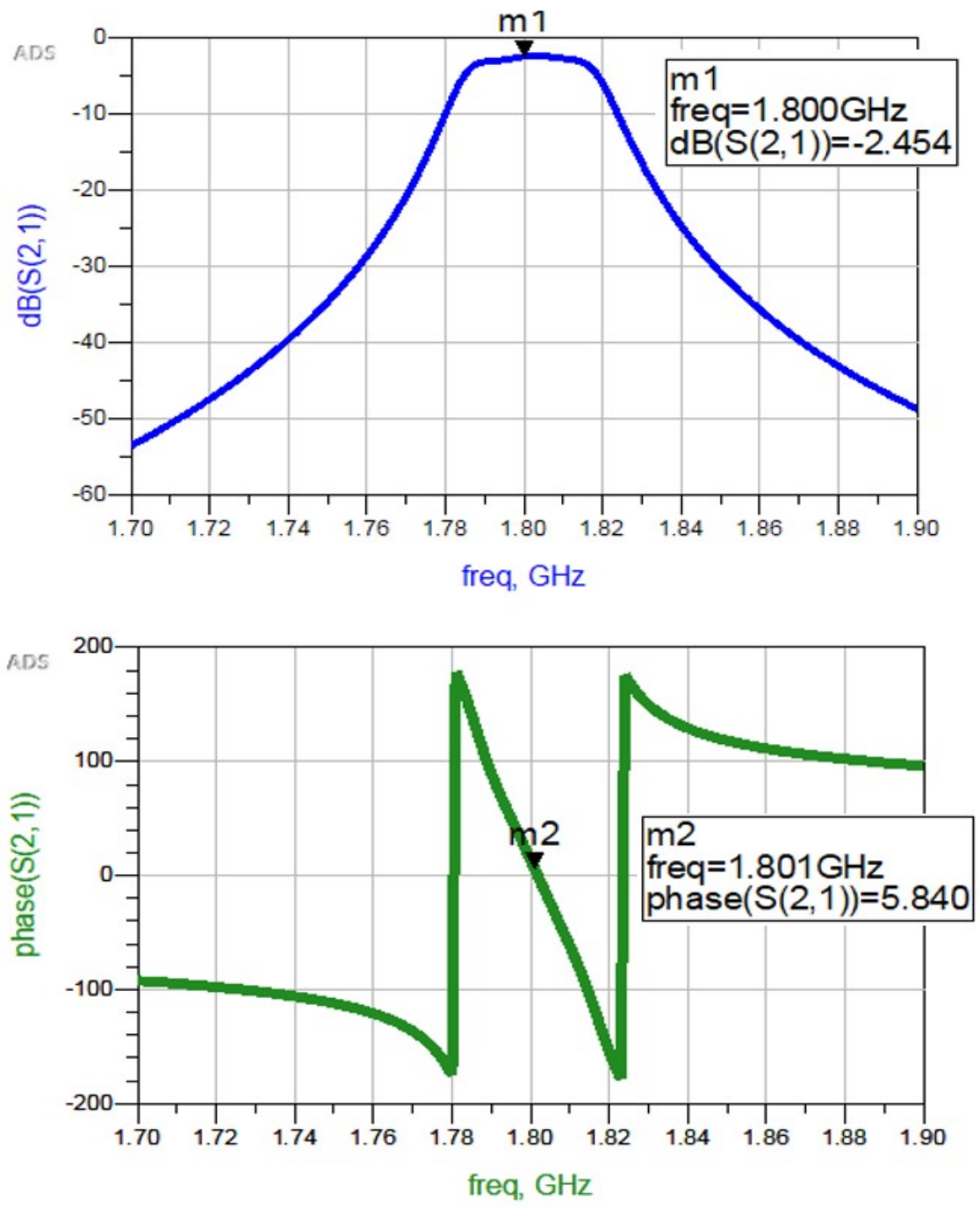

**Fig. 30.** Output of S-parameter analysis for the microstrip filter, as depicted in Figure 29.

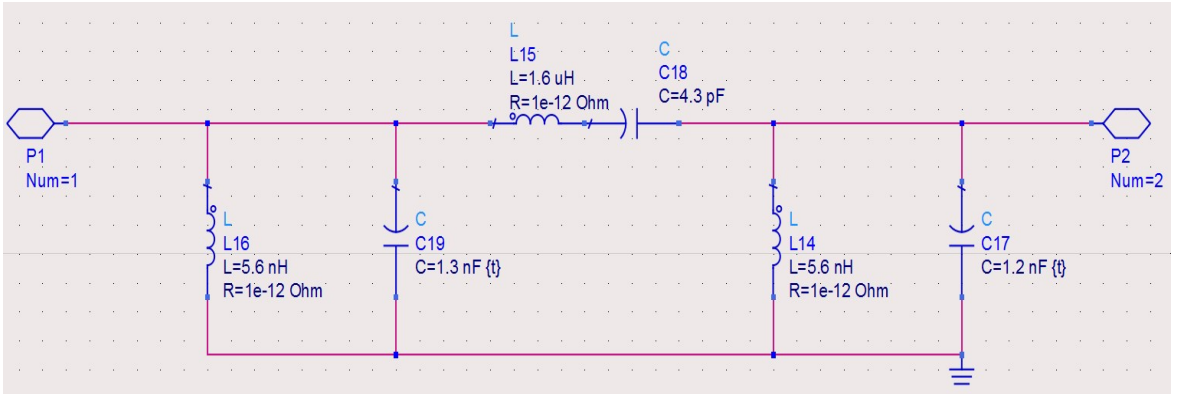

**Fig. 31.** LC filter placed through block titled I__6 in the RF chain of Figure 28 within ADS.

Following the simulation of the schematic in Figure 28, the results are presented in Table 12.

**Table 12.** Output of Link Budget simulation in Figure 28.

| System_Name | Label | System_Value |
|---|---|---|
| SystemInN0_dBm | | -173.98 |
| SystemInNPwr_dBm | | -173.98 |
| SystemInP1dB_dBm | P1dB | -32.48 |
| SystemInSOI_dBm | | 4.87 |
| SystemInTOI_dBm | IIP3 | -13.52 |
| SystemNF_dB | NF | 2.49 |
| SystemOutN0_dBm | | -121.06 |
| SystemOutNPwr_dBm | | -121.06 |
| SystemOutP1dB_dBm | OIP1 | 16.93 |
| SystemOutSOI_dBm | | 55.29 |
| SystemOutTOI_dBm | OIP3 | 36.90 |
| SystemPGain_SS_dB | | 50.42 |
| SystemPGain_dB | Gain | 50.42 |
| SystemPOut_dBm | | 0.42 |
| SystemS11_dB | | -13.47 |
| SystemS11_mag | | 0.21 |
| SystemS11_phase | | -34.51 |
| SystemS12_dB | | -400.00 |
| SystemS12_mag | | 0.00 |
| SystemS12_phase | | 0.00 |
| SystemS21_dB | | 50.42 |
| SystemS21_mag | | 331.77 |
| SystemS21_phase | | -92.26 |
| SystemS22_dB | | -14.52 |
| SystemS22_mag | | 0.19 |
| SystemS22_phase | | 3.56 |

Comparing the outputs highlighted in the red box with the desired values in Table 12 reveals that almost all of the desired specifications from Table 3 are achieved in the Link Budget simulation output. The only exception is the dynamic range of the entire RF chain, which remains unknown. The following equation can be used to calculate the dynamic range [7]:

$$\begin{aligned} Dynamic\ Range &= IP1 - MDS \\ &= OP1 - (G\text{-}1) - MDS\ \ (\text{dBm}) \end{aligned} \tag{5}$$

In (5), IP1 corresponds to P1dB at the input of the RF chain, while OP1 represents the value of such a parameter at the chain's output. G denotes the overall gain of the receiver, and MDS refers to the minimum detectable signal, typically considered equivalent to the system's overall sensitivity, calculated as follows [2]:

$$\begin{aligned} MDS = &\text{-}174(\text{dBm}) + NF \\ &+ 10\log(BW) + SNR\min(\text{dB}) \end{aligned} \tag{6}$$

When designing the RF chain for radar receivers in (6), it can be assumed that $SNRmin = 0\text{dB}$ [7]. Also, the bandwidth of the chain is defined by the narrowest bandwidth among its components, specifically the bandwidth of the final section, which was mentioned as 5 MHz. Consequently, there would be:

$$MDS = \text{-}174(\text{dBm}) + NF + 10\log(5\times10^6) \tag{7}$$

Now, by calculating MDS and using the values of OP1 and G from (7) and Table 10, the dynamic range could be determined as follows:

$$NF \approx 2.5\,\text{dB},\ G \approx\ 54.4\,\text{dB},\ OP1 = 16.95\,\text{dBm} \tag{8}$$

$$\begin{aligned} MDS &= \text{-}174(\text{dBm}) + NF + 10\log(5\times10^6) \\ &= \text{-}174 + 2.5 + 10\log(5\times10^6) \approx \text{-}104.5\,\text{dBm} \end{aligned} \tag{9}$$

$$\begin{aligned} &Dynamic\ Range = \\ &OP1(\text{dBm}) - (G\text{-}1)(\text{dB}) - MDS(\text{dBm}) = \\ &\quad 16.95 - (54.4\text{-}1) - (\text{-}104.5) \\ &\quad = 68.05\,\text{dB} \approx 68\,\text{dB} \end{aligned} \tag{10}$$

Based on the calculated dynamic range in (8)-(10) and its comparison with the desired value, which should be at least 70 dB in Table 1, it is evident that the dynamic range is approximately 2 dB lower than the desired value. To enhance the dynamic range, the system's degree of freedom, particularly the VGA, can be utilized. By adjusting the VGA gain, the optimal dynamic range could be achieved. As shown in Table 12, the overall gain of the chain is 54.4 dB, which exceeds the desired value in Table 3 by 4 dB. Thus, reducing the VGA gain by 4 dB will decrease the overall gain of the chain to 50.4 dB, improving some of the chain's nonlinear characteristics. The output from the simulation in Figure 28, reflecting this adjustment, is presented in Table 13.

Table 13 shows a slight decrease in OP1, which has reached 16.93 dBm. The NF value remains unchanged, as anticipated, based on the Friis formula. Now, the dynamic range could be recalculated.

**Table 13.** Output of Link Budget simulation shown after reduction of VGA's gain.

| System_Name | System_Value |
|---|---|
| SystemInN0_dBm | -173.98 |
| SystemInNPwr_dBm | -173.98 |
| SystemInP1dB_dBm | -36.47 |
| SystemInSOI_dBm | 4.87 |
| SystemInTOI_dBm | -16.89 |
| SystemNF_dB | 2.49 |
| SystemOutN0_dBm | -117.06 |
| SystemOutNPwr_dBm | -117.06 |
| SystemOutP1dB_dBm | 16.95 |
| SystemOutSOI_dBm | 59.29 |
| SystemOutTOI_dBm | 37.53 |
| SystemPGain_SS_dB | 54.42 |
| SystemPGain_dB | 54.41 |
| SystemPOut_dBm | 4.41 |
| SystemS11_dB | -13.47 |
| SystemS11_mag | 0.21 |
| SystemS11_phase | -34.51 |
| SystemS12_dB | -400.00 |
| SystemS12_mag | 0.00 |
| SystemS12_phase | 0.00 |
| SystemS21_dB | 54.41 |
| SystemS21_mag | 525.59 |
| SystemS21_phase | -92.26 |
| SystemS22_dB | -13.98 |
| SystemS22_mag | 0.20 |
| SystemS22_phase | 5.30 |

$$NF \approx 2.5\,\text{dB},\ G \approx\ 50.4\,\text{dB},\ OP1 = 16.93\,\text{dBm} \tag{11}$$

$$MDS = \text{-}174 + 2.5 + 10\log(5\times 10^{6}) \approx \text{-}104.5\,\text{dBm} \tag{12}$$

$$\begin{aligned} &\textit{Dynamic Range} = \\ &OP1(\text{dBm}) - (G\text{-}1)(\text{dB}) - MDS(\text{dBm}) = \\ &\quad 16.93 - (50.4\text{-}1) - (\text{-}104.5) \\ &\quad = 72.03\,\text{dB} \approx 72\,\text{dB} \end{aligned} \tag{13}$$

The new simulation's dynamic range also achieved the desired value, calculated through (11), (12), and (13). This result indicates that by delivering the suggested schematic, determining the frequency plan, and designing with the selected microwave commercial components discussed in section 3, the desired specifications outlined in Table 3 have been successfully met.

## 5. Conclusion

The reliable design of the RF chain is crucial for enhancing the performance of radar receivers, particularly in civil applications across various frequency bands. This paper highlights the importance of addressing image rejection using the advantages of intermediate frequencies. When removing the necessity of complex image-rejection blocks, the cost and complexity are reduced significantly, alongside improved attenuation of the image signal. However, the increased presence of spurious signals due to multiple mixers poses a challenge that cannot be overlooked, as it can severely limit the system's dynamic range. The scheme proposed to identify spurious-free regions for the frequency plan, developed in MATLAB using manual coding and the MATLAB toolbox, offers a robust solution for mitigating or surpassing these detrimental effects, facilitating a more efficient design process. The successful verification of this approach, after selecting and designing all components of the RF chain, combined with comprehensive evaluations using CASCADE software and the link budget analysis in ADS, confirms the viability of the proposed method. Besides, two detailed analytical designs of coupled-line filters that utilized full-wave approaches were presented and compared, along with their subsequent analysis, using the CST Studio GUI and the Layout section of ADS. This comparison provides practical insights into the design and implementation of microwave filters with intermediate frequency specifications. Overall, this work contributes to advancing RF chain design, paving the way for more effective radar receiver systems.

## 6. Future Work

While this paper has made significant progress in enhancing the design of RF chains for radar receivers and establishing a scheme for finding an appropriate frequency plan, there are several roads ahead for future work, including modification of design considerations. We can modify several specifications and considerations to evaluate the comprehensiveness of the proposed design method. These modifications could involve:

- Applying the proposed frequency plan design method to different frequency ranges.
- Testing the method on architectures with more mixers.
- Enhancing the dynamic range, then redoing the design process,
- Utilizing the proposed method to determine the frequency plan in up-conversion mode within transmitter mixers and the design of its RF chains.

## Acknowledgements

This work was supported by the Amirkabir Univeristy of Technology.

## References

[1] B. Razavi, *RF Microelectronics*, 2nd ed. Upper Saddle River, NJ, USA: Prentice Hall, 2011.

[2] Y. J. Ko, S. P. Stapleton, and R. Sobot, "Ku-Band Image Rejection Sliding-IF Transmitter in 0.13-μm CMOS Process," *IEEE Trans. Microwave Theory Tech., vol. 59, no. 8, pp. 2017-2019, Aug. 2011.*

[3] Y. Qi, Y. Wu, Q. Wang, W. Wang, and Z. Bai, "A tunable ultra-wideband superheterodyne radio frequency receiver with high-image-rejection levels," *Int. J. RF Microwave Computer. -Aided Eng., vol. 32, no. 1, 2022.*

[4] Z. Yu, J. Zhou, L. Zhao, and L. Dai, "The design of a 6GHz band RF receiver with excellent I/Q image suppression," *in 2012 International Conference on Microwave and Millimeter Wave Technology (ICMMT), vol. 4, pp. 1-4, May 2012. IEEE.*

[5] H. K. Yilmaz, H. Caylak, and S. Topaloglu, "Design of a Hartley Image-reject Receiver with Improved Image Rejection and Spurious-Free Frequency Synthesizer Response," *in 2019 IEEE Asia-Pacific Microwave Conference (APMC), pp. 51-53, Dec. 2019. IEEE.*

[6] G. F. De Andrade, E. V. S. Barbosa, and S. R. Rondineau, "RF Front-End Receiver for Vehicular Satellite Communications and LNA GaAs FET Design in Ku-Band," *in 2020 Workshop on Communication Networks and Power Systems (WCNPS), pp. 1-6, Nov. 2020. IEEE.*

[7] C. Budge and S. R. German, *Basic Radar Analysis*, Artech House Radar Series, Norwood, MA, USA: Artech House Publishers, 2015.

[8] M. I. Skolnik, *Radar Handbook*, 3rd ed. New York, NY, USA: McGraw-Hill Professional, 2008.

[9] D. Faria, L. Dunleavy, and T. Svensen, "The Use of Intermodulation Tables for Mixer Simulations," *Microwave Journal, vol. 45, no. 4, pp. 60, Dec. 2002.*

[10] Mini-Circuits, *MCA1-60+ Frequency Mixer Datasheet*, [Online]. Available: www.minicircuits.com/MCA-60+.

[11] Mini-Circuits, *ADE-35MH+ Frequency Mixer Datasheet*, [Online]. Available: www.minicircuits.com/ADE-MH35+.

[12] Mini-Circuits, *TSS-53LNB+ Wideband Low Noise Bypass Amplifier Datasheet*, [Online]. Available: www.minicircuits.com/TSS-53LNB+.

[13] Mini-Circuits, *GALI-74+ Monolithic Amplifier Datasheet*, [Online]. Available: www.minicircuits.com/GALI-74+.

[14] WJ Communications, Inc., *AH201 Medium Power, High Linearity Amplifier Product Datasheet*, [Online]. Available: datasheet4u.com/datasheet/ETC/AH201-136301.

[15] Analog Devices, *ADL5330 Voltage-Controlled Variable Gain Amplifier Datasheet*, [Online]. Available: www.analog.com/ADL5330.

[16] Mini-Circuits, *BFCN-2900+ Datasheet: Ceramic Bandpass Filter Datasheet*, [Online]. Available: www.minicircuits.com/BFCN-2900+.

[17] D. M. Pozar, *Microwave Engineering*, 4th ed. Hoboken, NJ, USA: Wiley, 2011.

[18] A. Ghorbani, G. Moradi, and S. M. Nargesi, *Advanced Microwave Engineering*, 1st ed. Iran: Niaze Danesh, 2014.